\documentclass[iop,twocolappendix,numberedappendix,apj]{emulateapj}
\usepackage{epsfig, natbib, graphicx, color, amsmath,
  cancel,icomma,bm,eso-pic,appendix,etoolbox}

\AddToShipoutPictureBG*{%
  \AtPageUpperLeft{%
    \hspace{0.75\paperwidth}%
    \raisebox{-8.0\baselineskip}{%
      \makebox[0pt][l]{\textnormal{DES 2016-0190}}
}}}%

\AddToShipoutPictureBG*{%
  \AtPageUpperLeft{%
    \hspace{0.75\paperwidth}%
    \raisebox{-9.0\baselineskip}{%
      \makebox[0pt][l]{\textnormal{SLAC PUB-16987}}
}}}%

\AddToShipoutPictureBG*{%
  \AtPageUpperLeft{%
    \hspace{0.75\paperwidth}%
    \raisebox{-10.0\baselineskip}{%
      \makebox[0pt][l]{\textnormal{Fermilab PUB-17-179-PPD}}
}}}%

\newcommand{\STD}{\mathrm{std}}
\newcommand{\alti}{\mathrm{alt}}
\newcommand{\az}{\mathrm{az}}
\newcommand{\obs}{\mathrm{obs}}
\newcommand{\ADU}{\mathrm{ADU}}
\newcommand{\AB}{\mathrm{AB}}
\newcommand{\ZPT}{\mathrm{ZPT}}
\newcommand{\ZPTAB}{\ZPT^\AB}
\newcommand{\inst}{\mathrm{inst}}
\newcommand{\atm}{\mathrm{atm}}

\newcommand{\DECal}{\mathrm{DECal}}
\newcommand{\flatt}{\mathrm{flat}}
\newcommand{\superstar}{\mathrm{superstar}}
\newcommand{\starflat}{\mathrm{starflat}}
\newcommand{\optics}{\mathrm{optics}}
\newcommand{\ccd}{\mathrm{ccd}}
\newcommand{\epoch}{\mathrm{epoch}}
\newcommand{\mjd}{\mathrm{MJD}}
\newcommand{\pixel}{\mathrm{pixel}}
\newcommand{\phot}{\mathrm{phot}}
\newcommand{\gray}{\mathrm{gray}}
\newcommand{\CCD}{\mathrm{CCD}}
\newcommand{\EXP}{\mathrm{EXP}}
\newcommand{\VAR}{\mathrm{VAR}}
\newcommand{\AUX}{\mathrm{AUX}}
\newcommand{\Sblong}{S_b(x,y,\alti,\az,t,\lambda)}
\newcommand{\Sbobs}{S_b^\obs(\lambda)}
\newcommand{\Sbstd}{S_b^\STD(\lambda)}
\newcommand{\PWV}{\mathrm{PWV}}
\newcommand{\pwv}{\mathrm{pwv}}
\newcommand{\pwvs}{\mathrm{pwv}_s}

\newcommand{\Fprime}{\mathcal{F}^{'}_\nu(\lambda_b)}
\newcommand{\Fprimeraw}{\mathcal{F}^{'}_\nu}
\newcommand{\Iten}{\mathbb{I}_{10}}
\newcommand{\Inaught}{\mathbb{I}_0}
\newcommand{\Ione}{\mathbb{I}_1}
\newcommand{\Oxygen}{\mathrm{O}_2}
\newcommand{\Ozone}{\mathrm{O}_3}
\newcommand{\FGCM}{\mathrm{FGCM}}
\newcommand{\fflag}{\mathrm{FLAG}^\FGCM}

\ifcsmacro{widetext}{}{
  \newenvironment*{widetext}{}{}
}

\def \figscalesmall {1.1}
\def \figscalebig {1.2}

\shorttitle{Forward Global Photometric Calibration of the Dark Energy Survey}  
\shortauthors{Burke et al.}

\begin{document}
\title{Forward Global Photometric Calibration of the Dark Energy Survey}

\def\andname{}

\author{
D.~L.~Burke\altaffilmark{1,2},
E.~S.~Rykoff\altaffilmark{1,2},
S.~Allam\altaffilmark{3},
J.~Annis\altaffilmark{3},
K.~Bechtol\altaffilmark{4},
G.~M.~Bernstein\altaffilmark{5},
A.~Drlica-Wagner\altaffilmark{3},
D.~A.~Finley\altaffilmark{3},
R.~A.~Gruendl\altaffilmark{6,7},
D.~J.~James\altaffilmark{8,9},
S.~Kent\altaffilmark{3,10},
R.~Kessler\altaffilmark{10},
S.~Kuhlmann\altaffilmark{11},
J.~Lasker\altaffilmark{10},
T.~S.~Li\altaffilmark{3},
D.~Scolnic\altaffilmark{10},
J.~Smith\altaffilmark{12},
D.~L.~Tucker\altaffilmark{3},
W.~Wester\altaffilmark{3},
B.~Yanny\altaffilmark{3},
T. M. C.~Abbott\altaffilmark{9},
F.~B.~Abdalla\altaffilmark{13,14},
A.~Benoit-L{\'e}vy\altaffilmark{15,13,16},
E.~Bertin\altaffilmark{15,16},
A. Carnero Rosell\altaffilmark{17,18},
M.~Carrasco~Kind\altaffilmark{6,7},
J.~Carretero\altaffilmark{19},
C.~E.~Cunha\altaffilmark{1},
C.~B.~D'Andrea\altaffilmark{5},
L.~N.~da Costa\altaffilmark{17,18},
S.~Desai\altaffilmark{20},
H.~T.~Diehl\altaffilmark{3},
P.~Doel\altaffilmark{13},
J.~Estrada\altaffilmark{3},
J.~Garc\'ia-Bellido\altaffilmark{21},
D.~Gruen\altaffilmark{1,2},
G.~Gutierrez\altaffilmark{3},
K.~Honscheid\altaffilmark{22,23},
K.~Kuehn\altaffilmark{24},
N.~Kuropatkin\altaffilmark{3},
M.~A.~G.~Maia\altaffilmark{17,18},
M.~March\altaffilmark{5},
J.~L.~Marshall\altaffilmark{25},
P.~Melchior\altaffilmark{26},
F.~Menanteau\altaffilmark{6,7},
R.~Miquel\altaffilmark{27,19},
A.~A.~Plazas\altaffilmark{28},
M.~Sako\altaffilmark{5},
E.~Sanchez\altaffilmark{29},
V.~Scarpine\altaffilmark{3},
R.~Schindler\altaffilmark{2},
I.~Sevilla-Noarbe\altaffilmark{29},
M.~Smith\altaffilmark{30},
R.~C.~Smith\altaffilmark{9},
M.~Soares-Santos\altaffilmark{3},
F.~Sobreira\altaffilmark{31,17},
E.~Suchyta\altaffilmark{32},
G.~Tarle\altaffilmark{33},
A.~R.~Walker\altaffilmark{9}
\\ \vspace{0.2cm} (DES Collaboration) \\
}

\affil{$^{1}$ Kavli Institute for Particle Astrophysics \& Cosmology, P. O. Box 2450, Stanford University, Stanford, CA 94305, USA}
\affil{$^{2}$ SLAC National Accelerator Laboratory, Menlo Park, CA 94025, USA}
\affil{$^{3}$ Fermi National Accelerator Laboratory, P. O. Box 500, Batavia, IL 60510, USA}
\affil{$^{4}$ LSST, 933 North Cherry Avenue, Tucson, AZ 85721, USA}
\affil{$^{5}$ Department of Physics and Astronomy, University of Pennsylvania, Philadelphia, PA 19104, USA}
\affil{$^{6}$ Department of Astronomy, University of Illinois, 1002 W. Green Street, Urbana, IL 61801, USA}
\affil{$^{7}$ National Center for Supercomputing Applications, 1205 West Clark St., Urbana, IL 61801, USA}
\affil{$^{8}$ Astronomy Department, University of Washington, Box 351580, Seattle, WA 98195, USA}
\affil{$^{9}$ Cerro Tololo Inter-American Observatory, National Optical Astronomy Observatory, Casilla 603, La Serena, Chile}
\affil{$^{10}$ Kavli Institute for Cosmological Physics, University of Chicago, Chicago, IL 60637, USA}
\affil{$^{11}$ Argonne National Laboratory, 9700 South Cass Avenue, Lemont, IL 60439, USA}
\affil{$^{12}$ Austin Peay State University, Clarksville, TN 37044, USA}
\affil{$^{13}$ Department of Physics \& Astronomy, University College London, Gower Street, London, WC1E 6BT, UK}
\affil{$^{14}$ Department of Physics and Electronics, Rhodes University, PO Box 94, Grahamstown, 6140, South Africa}
\affil{$^{15}$ CNRS, UMR 7095, Institut d'Astrophysique de Paris, F-75014, Paris, France}
\affil{$^{16}$ Sorbonne Universit\'es, UPMC Univ Paris 06, UMR 7095, Institut d'Astrophysique de Paris, F-75014, Paris, France}
\affil{$^{17}$ Laborat\'orio Interinstitucional de e-Astronomia - LIneA, Rua Gal. Jos\'e Cristino 77, Rio de Janeiro, RJ - 20921-400, Brazil}
\affil{$^{18}$ Observat\'orio Nacional, Rua Gal. Jos\'e Cristino 77, Rio de Janeiro, RJ - 20921-400, Brazil}
\affil{$^{19}$ Institut de F\'{\i}sica d'Altes Energies (IFAE), The Barcelona Institute of Science and Technology, Campus UAB, 08193 Bellaterra (Barcelona) Spain}
\affil{$^{20}$ Department of Physics, IIT Hyderabad, Kandi, Telangana 502285, India}
\affil{$^{21}$ Instituto de Fisica Teorica UAM/CSIC, Universidad Autonoma de Madrid, 28049 Madrid, Spain}
\affil{$^{22}$ Center for Cosmology and Astro-Particle Physics, The Ohio State University, Columbus, OH 43210, USA}
\affil{$^{23}$ Department of Physics, The Ohio State University, Columbus, OH 43210, USA}
\affil{$^{24}$ Australian Astronomical Observatory, North Ryde, NSW 2113, Australia}
\affil{$^{25}$ George P. and Cynthia Woods Mitchell Institute for Fundamental Physics and Astronomy, and Department of Physics and Astronomy, Texas A\&M University, College Station, TX 77843,  USA}
\affil{$^{26}$ Department of Astrophysical Sciences, Princeton University, Peyton Hall, Princeton, NJ 08544, USA}
\affil{$^{27}$ Instituci\'o Catalana de Recerca i Estudis Avan\c{c}ats, E-08010 Barcelona, Spain}
\affil{$^{28}$ Jet Propulsion Laboratory, California Institute of Technology, 4800 Oak Grove Dr., Pasadena, CA 91109, USA}
\affil{$^{29}$ Centro de Investigaciones Energ\'eticas, Medioambientales y Tecnol\'ogicas (CIEMAT), Madrid, Spain}
\affil{$^{30}$ School of Physics and Astronomy, University of Southampton,  Southampton, SO17 1BJ, UK}
\affil{$^{31}$ Instituto de F\'isica Gleb Wataghin, Universidade Estadual de Campinas, 13083-859, Campinas, SP, Brazil}
\affil{$^{32}$ Computer Science and Mathematics Division, Oak Ridge National Laboratory, Oak Ridge, TN 37831}
\affil{$^{33}$ Department of Physics, University of Michigan, Ann Arbor, MI 48109, USA}

\email{daveb@slac.stanford.edu}

\begin{abstract}
Many scientific goals for the Dark Energy Survey (DES)
require calibration of optical/NIR broadband $b = grizY$ photometry that is stable in time and
uniform over the celestial sky to one percent or better.
It is also necessary to limit to similar accuracy systematic uncertainty in the calibrated broadband magnitudes due to 
uncertainty in the spectrum of the source. Here we present a ``Forward Global Calibration Method (FGCM)'' for photometric calibration of the DES,
and we present results of its application to the first three years of the
survey (Y3A1).
The FGCM combines data taken with auxiliary instrumentation at the observatory
with data from the broad-band survey imaging itself and models of the instrument and atmosphere
to estimate the spatial- and time-dependence of the passbands of individual DES survey exposures.
``Standard'' passbands are chosen that are typical of the passbands encountered during the survey.
The passband of any individual observation is combined with an estimate of the source spectral shape
to yield a magnitude $m_b^{\STD}$ in the standard system. 
This ``chromatic correction'' to the standard system is necessary to achieve sub-percent calibrations.
The FGCM achieves reproducible and stable photometric calibration of standard magnitudes $m_b^{\STD}$ of stellar sources
over the multi-year Y3A1 data sample with residual random calibration
errors of $\sigma=5-6\,\mathrm{mmag}$ per exposure.
The accuracy of the calibration is uniform across the $5000\,\mathrm{deg}^2$ DES footprint to within $\sigma=7\,\mathrm{mmag}$.
The systematic uncertainties of magnitudes in the standard system due to the spectra of sources
are less than $5\,\mathrm{mmag}$ for main sequence stars with $0.5<g-i<3.0$.
\end{abstract}

\keywords {methods:observational - surveys - techniques:photometric}

\section{Introduction}
\label{Sec:introduction}

We present a ``Forward Global Calibration Method (FGCM)'' for photometric
calibration of ground-based wide-band optical/near-IR surveys such as the Dark Energy Survey
(DES)~\citep{des16} and the survey that will be carried out with the Large Synoptic Survey Telescope~\citep[LSST:][]{lsst09}.
We have applied this method to the first three years of the DES campaign~\citep{diehl16},
and achieve sub-percent reproducibility and uniformity in the multi-band photometry of this dataset.
This method also provides sufficiently detailed knowledge of the shape of the passband
of each survey exposure to account with similar precision
for the dependence of the photometry on the Spectral Energy Distribution (SED) of the source.
This ``chromatic correction'' is required for proper scientific interpretation
of the observed wide-band optical flux.

The FGCM is a photometric model-based approach~\citep{sandt06, burke10} to calibration of multi-band imaging surveys.
The FGCM does not rely on previously established ``standard stars'' or other celestial targets.
In-situ instrumentation is used
to periodically measure the optical properties of the survey instrumental
system, and additional dedicated equipment is used to continuously monitor
atmospheric conditions during periods of survey operations.  These auxiliary
data are combined with the repeated observations of stars found in the survey
data to ``forward'' compute the fraction of photons in the telescope beam at the top of the
atmosphere that are predicted to be detected in the sensors of the camera.
The FGCM iteratively solves for parameters of the photometric model that best fit
the number of photons observed in the camera.

The FGCM determination of passband ``throughput'' differs from other techniques that have been used to calibrate wide-field
surveys~\citep[e.g.][]{glazebrook94, macdonald04, padman08, regnault09, schlafly12, magnier16}.
These earlier works incorporate ``ubercal'' matrix formulations to obtain relative photometric normalizations for each exposure.
But they do not fully describe the passbands through which exposures are taken,  
nor do they account for the shapes of the SEDs of the calibration sources.
This leaves ambiguity in the scientific interpretation of the observed broad-band flux that can dominate the
measurement uncertainties.  
The FGCM selects ``photometric'' survey exposures that best sample the  
time-dependent atmospheric and instrumental passbands through which the survey is conducted.
It combines these with auxiliary data and photometric models to provide continuous calibration of the survey observing conditions.
This approach provides the shapes of the observing passbands as well as their relative normalizations.  
The goal is to convert a broad-band photometric measurement taken on any part of the focal plane at any time
of the multi-year survey to the value it would have in an invariant reference passband.

The instrumental response of modern ccd-based survey instruments
can vary continuously, but significant variations occur only over periods of days to weeks.
For our purposes, the Earth's atmosphere can be characterized by a small set of constituents that must
be tracked continuously throughout each night~\citep{stubbs07}.
Computation of the transmission of light from the top of the atmosphere (TOA) to the earth's
surface over a wide variety of these conditions 
can be done with extremely good accuracy with modern, and readily available,
computer programs~(\citealp[MODTRAN:][]{MODTRAN99}; \citealp[libRadTran:][]{libTran05}).

The FGCM is a two-step process.  First, parameters that define the instrumental
and atmospheric conditions during survey operations are fit to the broadband
survey data to establish an extensive network of calibration stars that spans
the survey footprint. The FGCM fit minimizes the dispersions of the repeated
measurements of fully corrected standard magnitudes of the calibration stars.
In this step the FGCM process identifies those exposures that allow best
extraction of the observing conditions during a given night.  The magnitudes of
the calibration stars are not explicit free parameters of the fit, but rather
computed from observed flux counts and the fitted photometric model parameters.
Therefore, the FGCM yields an extremely efficient parameterization of the
photometric calibration of the entire survey.  In the second step, the
calibration stars are used to determine observing conditions for individual
science exposures.  This calibration step does not require the exposures to
have been used in the first step nor taken in ``photometric'' conditions,
provided they have sufficient overlap with the calibration stars.

Features of the FGCM are:
\begin{enumerate}
\item The instrumental response and the make-up of the atmosphere can be characterized at
any time by a relatively small set of parameters.
\item These parameters vary in time slowly compared to the rate at which survey exposures are acquired.
\item We are free to choose survey exposures that best determine the calibration parameters and magnitudes of the calibration stars.
\item Data taken in any band will contribute to the calibration of all bands taken in the same period of time.
\item The FGCM incorporates data from auxiliary instrumentation when they are available;
it remains robust, though less precise, when auxiliary data are unavailable.
\item The FGCM is sensitive to the shape of the observing passband and allows correction for variation of the
Spectral Energy Distributions (SEDs) of celestial sources \citep{li16}.
\end{enumerate}

The FGCM does not determine the absolute flux scales of the reference
passbands.  Absolute calibration may be achieved via HST
CALSPEC\footnote{http://www.stsci.edu/hst/observatory/crds/calspec.html}
standards~\citep{bohlin07}, several of which are included in the DES footprint.
One of these (C26202) is within the footprint and dynamic range of normal DES
science exposures, so is particularly attractive as a possible source for
absolute calibration.  The possibility of using dedicated observations of stars
with nearly thermal SEDs (e.g. DA white-dwarfs) for standardization of color has
also been studied~\citep{smith15}.  However, absolute calibration is a topic
that is outside the scope of the work presented in this text.

The DES consists of repeated tilings of approximately $5000\,\mathrm{deg}^2$ of the Southern Sky in five wide-band filters $grizY$
with the Blanco telescope and DECam instrument~\citep{flaugher15} at the Cerro Tololo Inter-American Observatory (CTIO).
Approximately ten million relatively bright isolated stars are found in the DES footprint, and each will have been observed
in each of the five bands typically eight times at conclusion of the five-year survey.
Additional data are acquired with an in-situ multi-wavelength illumination ``DECal'' system \citep{marshall13} to measure
the wavelength dependence of transmission of light through the Blanco/DECam optical system (including the changeable filter)
and the spectral response of the sensors in the camera.
The DES also acquires real-time data from CTIO site meteorology instrumentation,
the SUOMINET GPS system\footnote{http://www.suominet.ucar.edu},
and auxiliary ``aTmCAM'' instrumentation~\citep{li14} to track changes in
conditions at the observatory and the make up of the atmosphere above the observatory.
The all-sky infrared cloud camera RASICAM~\citep{reil14} is used to guide observing operations.

The DES observations at CTIO are made over a combination of full and half nights equivalent to 105 full nights from August through February.
Initial ``Science Verification (SV)'' observations were made in the 2012-2013 season to commission the instrument and survey strategy,
and an ``SVA1 Gold'' data release\footnote{https://des.ncsa.illinois.edu/releases} is available for public use.
These data were calibrated with a version of the earlier ``ubercal'' technique,
and successfully met the DES design requirement of 2\% or better photometric accuracy~\citep{tucker07b}. 
The main survey began in August 2013, and the third of the planned five-year science campaigns was completed in February 2016.
We report here on the FGCM calibration of this first three-year ``Y3A1'' dataset.

In Section~\ref{sec: photometry} of this paper we first present the concept of
broad-band photometry with chromatic corrections, and follow in
Section~\ref{sec:FGCMform} with the formulation of the FGCM calibration model.
In Section~\ref{sec:FGCMproc}, we next discuss the FGCM process and the
execution of the calibration of the DES Y3A1 three-year data release.  In
Section~\ref{sec:y3a1perf} we define FGCM metrics and tests, and present
results of the performance of the FGCM calibration of Y3A1.  In
Section~\ref{sec:ZPT} we define and discuss the FGCM output data products and
their use.  Finally, in Section~\ref{sec:sumry} we discuss plans for further
improvements of the FGCM procedure.

\section{Broad-Band Photometry with Chromatic Corrections}
\label{sec: photometry}

A digital camera on a ground-based astronomical telescope will count a fraction of the photons produced by a
celestial source that reach the top of the earth's atmosphere.
For broad-band observations, the number of analog-to-digital counts (ADU) in the camera produced by a source
is proportional to the integral of the TOA flux $F_\nu(\lambda)$ from the
source weighted by the observational passband transmission, $\Sblong$ in broadband filter $b = \{grizY\}$:
\begin{equation}
\label{eqn:cnts}
\begin{split}
\ADU_b &= \frac{A}{g} \times \int_0^{\Delta T} {dt} \\
 & \qquad{} \times \int_0^\infty { F_\nu(\lambda) \times \Sblong \times
  \frac{d\lambda}{h_{Pl}\lambda} } ,
\end{split}
\end{equation}
where $A$ is the area of the telescope pupil, $g$ is the electronic gain of the camera sensors (electron/ADU),
and $\Delta T$ is the duration of the exposure.
The units of flux $F_{\nu}(\lambda)$ are $\mathrm{ergs}\,\mathrm{cm}^{-2}\,\mathrm{s}^{-1}\,\mathrm{Hz}^{-1}$,
and the factor $(h_{Pl}\lambda)^{-1}d\lambda$ counts the number of photons per unit energy at a given wavelength ($h_{Pl}$ is the Planck constant).
The coordinates $(x,y)$ are those of the source image in the focal plane of the camera,
$(\alti,\az)$ are the altitude and azimuth of the telescope pointing,
and $t$ is the time and date (modified Julian date; MJD) of the observation.  For convenience, we
refer to this position- and time-variable observational passband as:
\begin{equation}
\label{eqn:Sbobs}
  \Sbobs \equiv \Sblong.
\end{equation}

We define an observed TOA magnitude of a celestial source to be~ \citep{fukugita96},
\begin{equation}
\label{eqn:mtoa1}
m_b^{\obs}   \equiv - 2.5 \log_{10}\left( \frac{  \int_0^\infty  F_{\nu}(\lambda) \times \Sbobs \times \lambda^{-1} d\lambda }
           {  \int_0^\infty  F^\AB \times \Sbobs \times \lambda^{-1} d\lambda }         \right),
\end{equation}
where the AB flux normalization $F^{AB}$ = 3631 Jansky (1 Jy =
$10^{-23}\,\mathrm{ergs}\,\mathrm{cm}^{-2}\,\mathrm{s}^{-1}\,\mathrm{Hz}^{-1}$)\citep{OG83}.

With the measured ADU counts from Eqn. \ref{eqn:cnts}, this becomes,
\begin{equation}
\label{eqn:mtoa2}
\begin{split}
m_b^{\obs}  & = - 2.5 \log_{10}\left( \frac{ g \times \ADU_b }{A \times \Delta T \times F^{\AB} \times \int_0^\infty  \Sbobs \times (h_{Pl}\lambda)^{-1} d\lambda } \right)                                                           \\
         & = - 2.5 \log_{10} (\ADU_b) +  2.5 \log_{10} (\Delta T)                  \\
         & \qquad\qquad  {} + 2.5 \log_{10} \left(  \int_0^\infty \Sbobs
\times \lambda^{-1} d\lambda \right)  + \ZPTAB\\
         & = - 2.5 \log_{10} (\ADU_b) +  2.5 \log_{10} (\Delta T) \\
 & \qquad\qquad{} + 2.5 \log_{10} (\Inaught^\obs(b)) + \ZPTAB,
\end{split}
\end{equation}
where
\begin{equation}
\label{eqn:AB}
\ZPTAB = 2.5\log_{10} \left( \frac{A F^\AB}{g h_{Pl}} \right),
\end{equation}
and $\Inaught^\obs$ is defined as the integral over the observational
passband $b$:
\begin{equation}
\label{eqn:I0}
  \Inaught^\obs(b) \equiv \int_0^\infty \Sbobs \lambda^{-1} d\lambda.
\end{equation}

The utility of Eqns. \ref{eqn:mtoa1} and \ref{eqn:mtoa2} is limited by the large variety of passbands that will be
encountered during the course of the DES campaign.
Even if each passband is known,
proper scientific interpretation will depend on knowledge of the wavelength dependence of the source SED.
We seek to define a unique photometric quantity associated with each
source that can be compared to other measurements and theoretical predictions,
and we seek a method to obtain this quantity from the DES campaign data.

Consider the broad-band magnitude that would be measured if the source were observed through a ``standard'' passband $S^{\STD}_b(\lambda)$
that we choose at our convenience,
\begin{equation}
\label{eqn:mstd}
m_b^\STD   \equiv  -2.5 \log_{10}\left( \frac{  \int_0^\infty  F_{\nu}(\lambda) \times S^{\STD}_b(\lambda) \times \lambda^{-1} d\lambda }
           {  \int_0^\infty F^{\AB} \times \Sbstd \times \lambda^{-1} d\lambda }         \right).
\end{equation}  

\noindent The difference between this ``standard'' magnitude and a given observed magnitude is,
\begin{equation}
\label{eqn:STD-obs}
\begin{split}
\delta_b^\STD &\equiv m_b^\STD - m_b^\obs\\
  &= 2.5 \log_{10}( \Inaught^\STD(b) / \Inaught^\obs(b)) \\
  & \qquad{}  + 2.5 \log_{10}\left(
     \frac{ \int_0^\infty  F_{\nu}(\lambda) \times \Sbobs \times \lambda^{-1}
       d\lambda }
     {  \int_0^\infty F_{\nu}(\lambda) \times S^{\STD}_b(\lambda) \times
       \lambda^{-1} d\lambda } \right),
\end{split}
\end{equation}
where $\Inaught^{\STD}$ is defined analogously to Eqn.~\ref{eqn:I0} with the standard passband.
Given knowledge of the source SED and observational passband, this gives a unique transformation to
a magnitude in the corresponding standard passband.

In practice, direct use of Eqn.~\ref{eqn:STD-obs} is challenging.
We do not generally have detailed SEDs for all our photometrically-identified calibration stars,
and for the purposes of fitting model parameters, the amount of computing required to repeatedly perform the
necessary integrations is impractical.
However, as will be discussed in Section~\ref{sec:chromcorrs}, it is sufficient to utilize in the fit a
first-order expansion of the SED of each source that can be estimated from the observed colors of the star.
We write
\begin{equation}
\label{eqn:SEDexpand}
F_\nu(\lambda) = F_\nu(\lambda_b) + F^{'}_\nu(\lambda_b) (\lambda - \lambda_b),
\end{equation}
where
\begin{equation}
\label{eqn:lnSED}
F^{'}_\nu(\lambda) = \frac{dF_\nu(\lambda_b)}{d\lambda}
\end{equation}
is the average slope of the SED across the passband.
The prescription used by FGCM to compute suitably accurate SED slopes is given in Appendix \ref{app:SEDslope}.
For convenience, we additionally define the ratio
\begin{equation}
\label{eqn:Fratio}
\Fprime \equiv F^{'}_\nu(\lambda_b)/F_\nu(\lambda_b).
\end{equation}
The reference wavelength $\lambda_b$ is arbitrary; we define it as the
photon-weighted mean wavelength of the instrumental passband
\begin{equation}
  \label{eqn:lambdab}
\lambda_b \equiv \frac{\int_0^\infty \lambda \times S^\inst_b(\lambda) \times
\lambda^{-1} d\lambda}{\int_0^\infty S^\inst_b(\lambda) \times \lambda^{-1}
  d\lambda},
\end{equation}
where $S^\inst_b$ is the focal-plane average instrumental system response excluding the atmosphere.
With these definitions, Eqn.~\ref{eqn:STD-obs} becomes:
\begin{widetext}
\begin{equation}
\label{eqn:STD}
\delta_b^\STD \approx 2.5 \log_{10}( \Inaught^\STD / \Inaught^\obs) + 2.5
\log_{10}\left( \frac{ \int_0^\infty (1 + \Fprime \times (\lambda - \lambda_b))
  \times \Sbobs \times \lambda^{-1} d\lambda }{  \int_0^\infty (1 + \Fprime
  \times(\lambda - \lambda_b)) \times  \Sbstd \times \lambda^{-1} d\lambda } \right).
\end{equation}
\end{widetext}
We further define an $\Ione$ integral similar to Eqn.~\ref{eqn:I0},
\begin{equation}
\label{eqn:I1}
  \Ione^\obs(b) \equiv  \int_0^\infty {\Sbobs(\lambda - \lambda_b) \lambda^{-1} d\lambda},
\end{equation}
with a similar definition for the corresponding integral over the standard
passband.  It is also convenient to define the ``normalized chromatic passband integral''
\begin{equation}
\label{eqn:nci}
\Iten^\obs(b) \equiv \frac{\Ione^\obs(b)}{\Inaught^\obs(b)}.
\end{equation}
Note that in our linearized formulation $\Inaught$, $\Ione$, and $\Iten$ are all independent of the source SED.

Combining Eqn.~\ref{eqn:mtoa2} with Eqn.~\ref{eqn:STD} we obtain
\begin{equation}
\label{eqn:mstdfinal}
\begin{split}
  m_b^\STD = & -2.5 \log_{10} (\ADU) + 2.5 \log_{10}(\Delta T ) \\
  & \qquad{} + 2.5 \log_{10} (\Inaught^\obs) + \ZPTAB      \\ 
   & \qquad {} + 2.5 \log_{10} \left( \frac{ 1 + \Fprime \Iten^\obs(b)} {1 + \Fprime \Iten^\STD(b) } \right).
\end{split}
\end{equation}
The standard magnitude is determined by an ``instrumental magnitude'' given by raw ADU counts and exposure time,
a ``zero point'' integral of the observational passband with AB normalization, and a ``chromatic correction''.
Note that the chromatic correction will be zero if the observing
passband is the standard passband, so it is advantageous to choose standard
passbands that are those most often encountered during the survey.  The
correction will also be zero if the SED is flat across the passband, and the correction is unaffected by normalization of the passbands.

\section{Forward Global Calibration Formulation}
\label{sec:FGCMform}

The DES data management (DESDM) software package (Morganson, et al. 2017, in preparation) processes single DECam exposures using a dedicated version
of the well-known ``Source Extractor'' software \citep{bertin96} to produce ``FINALCUT'' catalogs (FITS databases)
of instrumental data from individual observations of celestial objects.
Corrections have been applied to these data for a number of instrumental effects including electronic bias and non-linearity,
variation in pixel-to-pixel response, and variation in the observing point-spread-function (including dependence on source brightness).
Sky backgrounds have been subtracted and images have been screened to remove those that exhibit a number of observing
or instrumental effects (Bernstein et al. 2017, in preparation). 
The FINALCUT catalogs are queried ({\it c.f.} Appendix \ref{app:init}) and the observational data
are processed with software that implements the FGCM photometric calibration.

The FGCM formulation follows the sequence in time over which the DES survey data are acquired.
The FGCM model parameters include some that are continuous functions of time, some that vary nightly,
some that vary over periods of months,
and others that change only when some ``event'' occurs such as instrumental maintenance.
The FGCM model does not include any {\it ad hoc} parameters unique to a given exposure.    
Assuming the instrumental properties do not depend on the atmospheric conditions,
the observational passband (Eqn.~\ref{eqn:cnts}) can be separated into two functions,  
\begin{equation}
\label{eqn:optb}
\begin{split}
\Sbobs & \equiv \Sblong \\
 & = S_b^\inst(x,y,t,\lambda) \times S^\atm(\alti,\az,t,\lambda),
\end{split}
\end{equation}
where $S^\atm$ is the transmittance (dimensionless) of photons from the top of the atmosphere to the input pupil of the telescope,
and $S_b^\inst$ is the response (CCD electrons/photon) of the instrumental system with optical filter $b$ to photons
that pass through the input pupil of the telescope.

\subsection{Instrumental System Response}

The response of the combined Blanco and DECam instrumental
system can be factored into parts that are characterized and determined in
different ways,
\begin{equation}
\label{eqn:Sinst}
\begin{split}
S_b^\inst(x,y,t,\lambda) & = S_b^\flatt(\pixel,\epoch) \times
S_b^\starflat(\pixel,\epoch) \\
& \qquad{} \times S_b^\superstar(\ccd,\epoch) \times S^\optics(\mjd)\\
& \qquad{} \times S_b^\DECal(\ccd,\lambda),
\end{split}
\end{equation}
where the independent variables (described in greater detail below) have been replaced with units that
appropriately match the granularity and stability of the system: pixels in each
CCD and MJD dates during an epoch of stable instrumental performance.
New epochs are defined at the start of yearly operations and whenever the
instrumental complement or performance of sensors is known to change.
There were five such epochs defined over the course of Y3A1.
None of the factors in Eq.~\ref{eqn:Sinst} include variations that might occur on hourly time scales
such as could be caused by instability in the temperatures of the sensors or electronics.
The average temperature of the DECam focal plane is maintained by an active thermal system,
and is found vary by no more than 0.1$^{\circ}$C over periods of weeks.
The response of the sensors over this range is expected to vary less than 0.1\%~\citep{estradaetal10},
and any such instability is included in the FGCM performance metrics discussed below.
We note also that only $S_b^\DECal$ has specific wavelength dependence,
and that this quantity includes nearly all of the loss of light through the system (Fig.~\ref{fig:Blanco/DECam_Passbands} below).  

\subsubsection{Flat Fields and Star Flats: $S_b^\flatt$ and $S_b^\starflat$}

Pixel-to-pixel variations in the detection efficiency of in-band
photons that pass through the telescope pupil are denoted by
$S_b^\flatt(\pixel,\epoch)$.  Electronic bias and traditional broadband
pixel-level ``flat'' frames for each filter band are obtained nightly,
and averages for each observing epoch are computed from a subset of the images for each CCD.
These are applied by DESDM to correct raw ADU counts during processing of science images.
This removes small spatial scale variations in sensor efficiency and variations in read-out amplifier gains.
However, this technique introduces well-known errors (see {\it e.g.}~\citep{regnault09} and references therein)
due to non-uniformity of the illumination pattern produced by the flat-field screen, and worsens
distortions of the projections of pixel shapes onto the celestial sky.

Dedicated exposures dithered across dense star fields are acquired once per observing epoch
and used by DESDM during processing of science images to correct for
large-scale non-uniformity in the instrument response left by the flat-field
process.  The starflat correction $S_b^\starflat(\pixel,\epoch)$ is
defined for each filter band on a sub-CCD spatial scale for each epoch.

The philosophy of the FGCM is to consider the acquisition and use of nightly broadband flats and dithered star corrections
by DESDM to be part of the overall system to be calibrated.
These pixel-level corrections are incorporated into the FINALCUT instrumental magnitudes that are input to the FGCM process. 

\subsubsection{Superstar Flats: $S_b^\superstar$}

The FGCM allows for refinement of the star flats which we refer to
as a ``superstar'' flat.  Denoted as $S_b^\superstar(\ccd,\epoch)$, this
correction is computed from the calibration exposures for each epoch for each
CCD at the end of each cycle of the calibration fit (Sec.~{\ref{sec:FGCMproc}).  In practice, this is
effectively a modification of the DESDM processing, and therefore we apply this
correction to the instrumental magnitudes before the next cycle of the fit.
This improves the accuracy and efficiency of the fitting process.
The superstar flats obtained for the Y3A1 calibration are discussed in
Section~\ref{sec:results:superstar} and examples shown in Figures~\ref{fig:superflatg}
and \ref{fig:superflati}.

\subsubsection{Opacity of Optical System: $S^\optics$}

The factor $S^\optics(\mjd)$ includes the opacity of the optical system created
by environmental dust that accumulates on the exposed optical surfaces.
It also includes the degradation in the reflectivity of the primary mirror bare aluminum
surface coating over time.  Dust is composed of particles with sizes large
compared with the wavelength of optical light, so its opacity is independent of
wavelength (i.e., gray).  It is optically located at the input pupil of the
telescope, so dust extinction is to good approximation independent of location
of the image in the focal plane.  The Blanco primary mirror is wet-washed several
times per year; a total of seven times during the DES Y3A1 three-year observing campaign.
It is also cleaned weekly with CO$_{2}$ gas.
The optical thickness of dust contamination is discontinuous
at each wash date and found to vary by several percent between wet washings.

The aluminized primary mirror was resurfaced in March 2011.
Engineering measurements show losses in reflectivity from the mirror of several percent per year
over the course of DES Y3A1 operations.
There is some evidence for wavelength dependence in the initial measurements,
but little in those taken following washes during DES operations.
No direct measurements are available of the absolute transparency of the exposed surface of the DECam entry window.
The FGCM model combines the effects of dust and mirror reflectivity into a single time-dependent wavelength-independent ``gray'' term
normalized to unity on a date near the start of Y3A1 observing.  The results
from the opacity fits are discussed in Section~\ref{sec:results:opacity} and
Figure~\ref{fig:FGCMoptics}.

\subsubsection{Wavelength Dependence from DECal: $S_b^\DECal$}

The in-situ ``DECal'' system
provides nearly monochromatic illumination of the DECam focal plane through the Blanco input pupil.
This system was used to measure the detailed wavelength dependencies $S_b^\DECal(\ccd,\lambda)$
of the $grizY$ instrumental passbands at the beginning of DES operations and once per year during the campaign.
These measurements are made with 2nm FWHM spectral bins stepped in 2nm increments across the nominal passband of each filter,
and in 10nm increments at wavelengths that are nominally ``out-of-band''
(defined as wavelengths approximately 10nm or more outside the main passband of the filter).
These data account for the wavelength dependence of the reflectivity of the primary mirror, the filter passbands, and the sensor efficiency.
The passbands are measured individually for each CCD in the DECam focal plane as shown in Figure~\ref{fig:Blanco/DECam_Passbands}.
The normalization is arbitrarily chosen to be the average of the CCD responses
over the central $\pm400~\mbox{\AA}$ of the $i$-band.  The light gray lines show the per-ccd
variation, which is especially pronounced for the $g$-band due to variations in
the quantum efficiencies of the sensors at the blue side.  In addition,
Figure~\ref{fig:ibandblueedge} shows the variation in the blue edge of the
$i$-band passband as a function of radius from the center of the
field-of-view; this is caused by variation of the transmittance of the filter with incidence angle.
The shapes of the passbands are measured with better than 0.1\% precision,
and are found to be stable over the Y3A1 campaign to the accuracy with which they are measured.

\begin{figure}
\scalebox{\figscalesmall}{\plotone{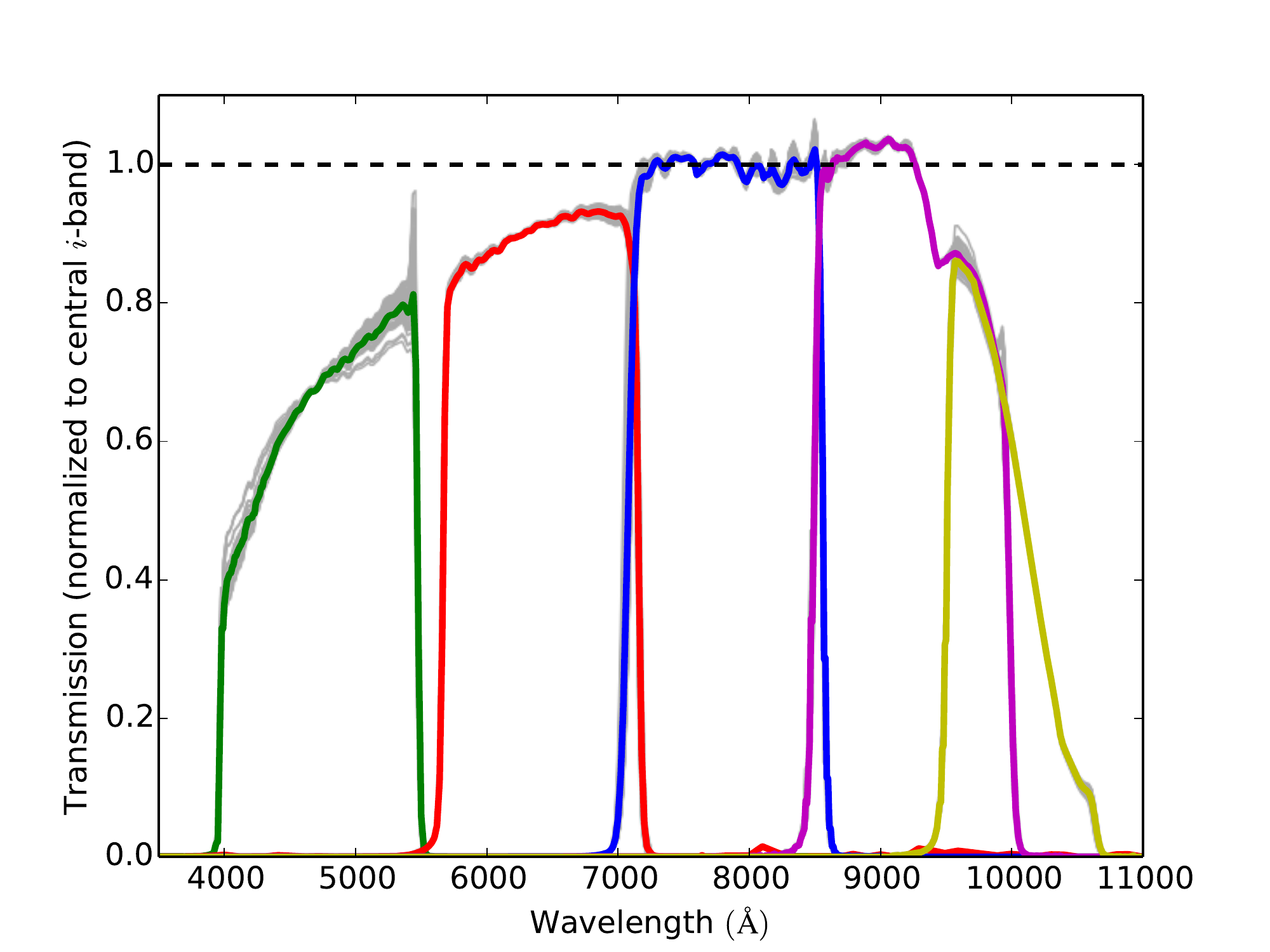}}
\caption{Blanco/DECam instrumental passbands $S_b^\DECal$ measured with the
  DECal system.  The solid color lines show the focal-plane average for
  $g$-band (green), $r$-band (red), $i$-band (blue), $z$-band (magenta), and
  $Y$-band (yellow).  In addition one light gray line plotted for each
  individual CCD shows the variation in response which is especially
  pronounce for the $g$-band.
  The variation of the blue edge of the $i$-band is shown in more detail in Figure~\ref{fig:ibandblueedge}.}
\label{fig:Blanco/DECam_Passbands}
\end{figure}

\begin{figure}
\scalebox{\figscalesmall}{\plotone{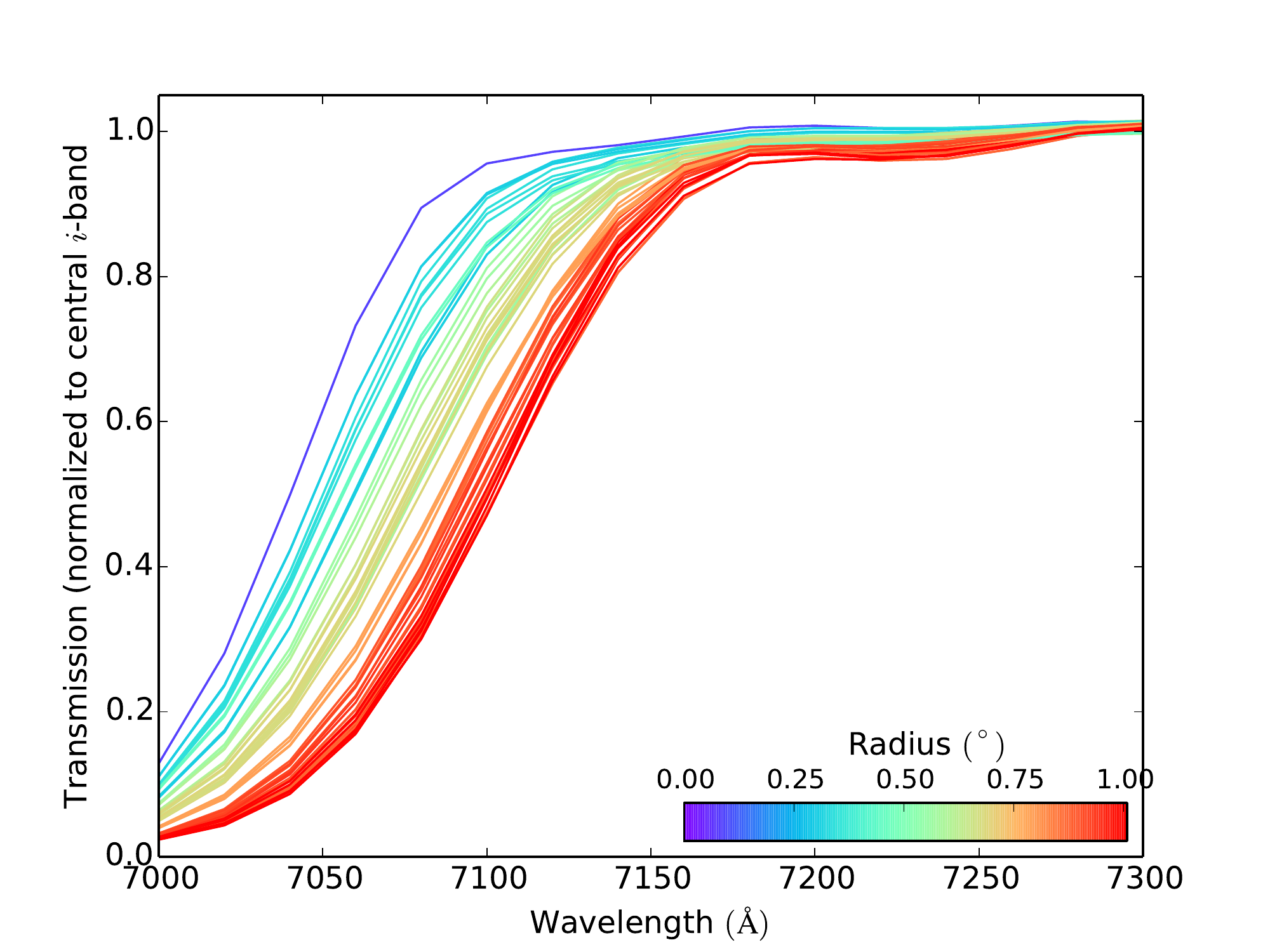}}
\caption{Radial variation of the blue edge of the $i$-band passband due to the
  filter.  Each line represents one CCD, and the color represents the distance (degrees on the sky)
  of the CCD from the center of the field-of-view.}
\label{fig:ibandblueedge}
\end{figure}

\subsubsection{Instrumental Fit Parameters}
\label{sec:instfit}

The vector of parameters of the instrumental system used to fit the observed DES data,
\begin{equation}
\label{eqn:FGCMinst}
\vec{P}^\inst \equiv ( \mathrm{optics}(\mathrm{wash}\_\mjd), \mathrm{rate}(\mathrm{wash}\_\mjd ) )
\end{equation}
includes the opacity $S^{\optics}(\mjd)$ of the optics after the primary mirror is washed on ${\rm wash}\_\mjd$,
and the linear rate of change in throughput of the optics during the period of time following each washing.
All other characteristics of the instrumental system are measured quantities. 

\subsection{Atmospheric Extinction}
\label{sec:atmext}

Processes that attenuate light as it propagates through the atmosphere include
absorption and scattering (Rayleigh) by molecular constituents ($\Oxygen$,
$\Ozone$, and trace elements), absorption by precipitable water vapor (PWV),
scattering (Mie) by airborne macroscopic particulate aerosols with physical
dimensions comparable to the wavelength of visible light, and shadowing by
larger ice crystals and water droplets in clouds that is independent of
wavelength (gray).

The FGCM fitting model for atmospheric transmittance is written,
\begin{equation}
\begin{split}
\label{eqn:Satm}
S^\atm(\alti,\az,t, \lambda) & = S^{\mathrm{molecular}}({\rm bp,zd,t},\lambda) \times S^\pwv({\rm zd,t},\lambda)
\\
& \qquad{}  \times e^{-({\rm X(zd)} \times \tau(t,\lambda))},
\end{split}
\end{equation}
where $S^{\mathrm{molecular}}$ accounts for absorption and scattering by dry gases, $S^\pwv$ accounts for
absorption by water vapor, $\tau$ is the aerosol optical depth, and $\mathrm{X}$ is the airmass.
The barometric pressure $\mathrm{bp(mm)}$, the zenith distance of the observation $\mathrm{zd}(\mathrm{deg})$,
and the time of the observation $t$ (MJD) are acquired for each exposure.
The airmass $X(\mathrm{zd})$ is computed separately for each exposure at the center of each CCD on the focal plane,
and includes corrections for the curvature of the earth \citep{kasten89} that become important at larger zenith distances.
As discussed below, the model does not explicitly include possible extinction by cloud cover.
We now describe in greater detail each of these terms and their corresponding parameterizations.

\subsubsection{Molecular Absorption: $S^{\mathrm{molecular}}$}
\label{sec:molecular}

Molecular Rayleigh scattering and absorption by oxygen and trace elements are
determined to high precision via the barometric pressure which is continuously
acquired as part of the environmental monitoring of the observing site.
Absorption of light by ozone can be characterized by one parameter; the integrated vertical column height
which varies seasonally with small variations over periods of days.
Therefore, we fit one parameter for ozone for each calibratable night.

\subsubsection{Water Vapor Absorption: $S^\pwv$}
\label{sec:pwv}

The FGCM parameterization of atmospheric transmission accounts for time
variations in precipitable water vapor during observing nights.  These are taken from
measurements made with auxiliary information when they are
available.  Auxiliary data are assigned to a DES survey exposure if their MJD
acquisition dates are most closely matched to, and are within $\pm2.4\,\mathrm{hours}$
of, that survey exposure.  The procedure adapts to missing auxiliary data by
inserting a model that is linear in time through the night. 

If a calibration exposure is successfully matched by auxiliary data, then the
precipitable water vapor is parameterized as
\begin{equation}
\label{eqn:pwvaux}
\PWV(\mathrm{exposure}) = \pwv_0(\mathrm{night}) + \pwv_1 \times \pwv_\AUX(\mathrm{exposure}),
\end{equation}
where $\pwv_\AUX$ is the value from the auxiliary instrumentation matched to
the exposure.  The nightly $\pwv_0$ term accounts for possible instrumental
calibration offsets in the auxiliary data, and the constant $\pwv_1$ (a single value for the entire run)
accommodates possible theoretical or computational scale differences between
the FGCM and auxiliary data reductions.
If the auxiliary instrument does not provide data for an exposure, then the
precipitable water vapor is parameterized with the less accurate approximation
\begin{equation}
\label{eqn:pwvut}
\PWV(\mathrm{exposure}) = \pwv(\mathrm{night}) + \pwvs(\mathrm{night}) \times \mathrm{UT}(\mathrm{exposure}),
\end{equation}
where the value at $\mathrm{UT} = 0$ ($\pwv(\mathrm{night})$) and time derivative
($\pwvs(\mathrm{night})$) are FGCM fit parameters.
The code allows for both cases within each night,
so requires three parameters per calibratable night plus the one overall scale parameter $\pwv_1$.

\subsubsection{Aerosol Absorption: $e^{-(\mathrm{X}\tau)}$}
\label{sec:aerosol}

Scattering by aerosols can be more complex, but the corresponding optical depth
for a single particulate species is well-described with two parameters as
\begin{equation}
\label{eqn:aertau}
\tau(\lambda) = \tau_{7750} \times (\lambda/7750\,\mbox{\AA})^{-\alpha}.
\end{equation}
The normalization $\tau_{7750}$ and optical index $\alpha$ depend on the
density, size, and shape of the aerosol particulate.
The FGCM does not use any of the available MODTRAN aerosol models, as these are specific to types of sites. 

Aerosol optical depth, like water vapor, can vary by several percent over
hours, so the calibration measurements and process must account for variations of this magnitude on these timescales. 
The aerosol normalization $\tau_{7750}$ is parameterized in a manner similar to the precipitable water vapor.
When auxiliary data are available
\begin{equation}
\label{eqn:tauaux}
\tau_{7750}(\mathrm{exposure}) = \tau_0(\mathrm{night}) + \tau_1 \times \tau_\AUX(\mathrm{exposure}),
\end{equation}
or if no auxiliary data are available
\begin{equation}
\label{eqn:tauut}
\tau_{7750}(\mathrm{exposure}) = \tau(\mathrm{night}) + \tau_s(\mathrm{night}) \times \mathrm{UT}(\mathrm{exposure}).
\end{equation}
Again, the code allows for both cases within each night,
so it requires three parameters per calibratable night plus the one overall scale parameter $\tau_1$.

For our present modeling, we assume that the aerosols on any given night are
dominated by a single species.  Therefore, we require one value for the aerosol
optical index $\alpha$(night) for each calibratable night.

\subsubsection{Atmospheric Fit Parameters}
\label{sec:atmofit}

The vector of atmospheric parameters used to fit the observed DES data,
\begin{equation}
\label{eqn:FGCMatm}
 \vec{P}^\atm \equiv (\Ozone, \pwv_0, \pwv_1, \pwv, \pwv_s, \tau_0, \tau_1, \tau, \tau_s, \alpha)
\end{equation}
includes the vertical column height of ozone (Dobson), the vertical column height of precipitable water vapor (mm),
the vertical optical depth of aerosol (dimensionless), and the aerosol optical index (dimensionless).

\subsection{Clouds, Photometric Conditions, and ``Gray'' Corrections}
\label{sec:clouds}

Observing operations for the DES are generally carried out only when the sky is relatively free of cloud cover.
Even so, condensation of water droplets and ice can produce thin clouds that are invisible to the naked eye
and have intricate spatial structure \citep[e.g.][]{burke14}.
This condensation process occurs along sharp boundaries in temperature and pressure determined by the volume density of 
precipitable water vapor;
this leads to the common characterization of observing conditions as either ``photometric'' or not.

The FGCM fitting model does not include a specific component for extinction by clouds,
but a rigorous procedure is followed to identify photometric, or nearly photometric, exposures for use in the calibration fit.
Estimates of the standard magnitudes $\overline{m_b^{\STD}}$ of the calibration stars obtained in each cycle of the FGCM fitting process
are used to estimate the extinction of each exposure that is not accounted for by the fitted parameter vectors.
(See Section~\ref{sec:FGCMfit} below for discussion of this process.)
This estimate is used to select the sample of calibration exposures to be used in the next cycle of the fitting process.  
During the DES observing season, the conditions at CTIO are such that cloud formation does not occur for large periods of time on many nights.
The FGCM finds that nearly 80\% of the exposures taken in the Y3A1 campaign were acquired under photometric conditions
and are used in the final fit cycle.

In the final step of the FGCM, an estimate of a ``gray'' correction is made from the observed $m_b^{\STD}$ on each exposure that accounts
for cloud extinction.
This step is discussed in detail in Sec.~\ref{sec:fgcmgrycor} below,
but we note here that this ``gray'' correction is an estimate of the cumulative effect from a number of sources that are not
explicit in the fitting model. 
This includes possible instrumental effects ({\it e.g.} dome occultations and shutter timing errors),
and residual errors in assignments of ADU counts to celestial sources ({\it e.g.} aperture corrections and subtraction of sky backgrounds). 
These may depend on the band of the exposure,
but are assumed to have no explicit wavelength dependence across each band, and so are labeled ``gray''.

\subsection{Standard Passbands and Observational Look-Up Tables}
\label{sec:stdsandluts}

Standard passbands were defined for the Y3A1 campaign, and look-up tables (LUTs) were pre-computed to allow
rapid evaluation of the passband integrals $\Inaught^\obs$ and $\Ione^\obs$ over a wide range of model parameter vectors. 

The standard instrumental system responses were chosen to be the average responses of the
CCDs in the focal plane shown in Figure \ref{fig:Blanco/DECam_Passbands}.
These were synthesized from DECal scans taken during the first two years of DECam operations.
The standard atmospheric transmittance was computed with the MODTRAN IV code~\citep{MODTRAN99} with the
parameters given in Table \ref{table:STDatmo} chosen as typical of those
encountered during the Y3A1 campaign.  The transmission of the various
components of the standard atmosphere are shown in
Figure~\ref{fig:standardAtm}, and the combined set of standard passbands are
shown in Figure \ref{fig:Standard_Passbands}.
Subsequent observations of the SDSS standard BD+17$^{\circ}$4708~\citep{fukugita96} indicate these passbands should be
multiplied by a factor $\approx0.55$ if approximate normalization is
desired.
Code that contains these passbands, as well as tools to use them are available for
download\footnote{https://opensource.ncsa.illinois.edu/bitbucket/\allowbreak projects/\allowbreak
  DESDM/repos/fgcm\_y3a1\_tools}.
The photon-weighted average wavelength, and the $\Inaught^\STD$ and $\Ione^\STD$ integrals and their ratio $\Iten^\STD$
for these passbands are given in Table \ref{table:STDparms}.

\begin{deluxetable}{lcr}
\tablewidth{0pt}
\tablecaption{Standard Atmosphere}
\tablehead{
  \colhead{Parameter} &
  \colhead{Units} &
  \colhead{Value}
}
\startdata
 Barometric Pressure          &   mb          &     778.0              \\
 Precipitable Water Vapor     &   mm          &     3.0                \\
 Ozone                        &   Dobson      &     263.0              \\
 Aerosol Optical Depth        &   None        &     0.030              \\
 Aerosol Optical Index        &   None        &     1.00               \\
 Airmass                      &   None        &     1.2                \\
\enddata
\label{table:STDatmo}
\end{deluxetable}

\begin{deluxetable}{crrrr}
\tablewidth{0pt}
\tablecaption{Standard Photometric Passband Parameters}
\tablehead{
  \colhead{Band} &
  \colhead{$\lambda_b$ (\AA)} &
  \colhead{$\Inaught^\STD$} &
  \colhead{$\Ione^\STD$ (\AA)} &
  \colhead{$\Iten^\STD$ (\AA)}
}
\startdata
 $g$      &   4766.0         &     0.163         &    4.333         &    26.58                    \\
 $r$      &   6406.1         &     0.187         &    1.850         &     9.89                    \\
 $i$      &   7794.9         &     0.174         &    1.344         &     7.72                    \\
 $z$      &   9174.4         &     0.136         &   -2.163         &   -15.90                    \\
 $Y$      &   9874.5         &     0.052         &    0.911         &    17.52                    \\
\enddata
\label{table:STDparms}
\end{deluxetable}

\begin{figure}
\scalebox{\figscalesmall}{\plotone{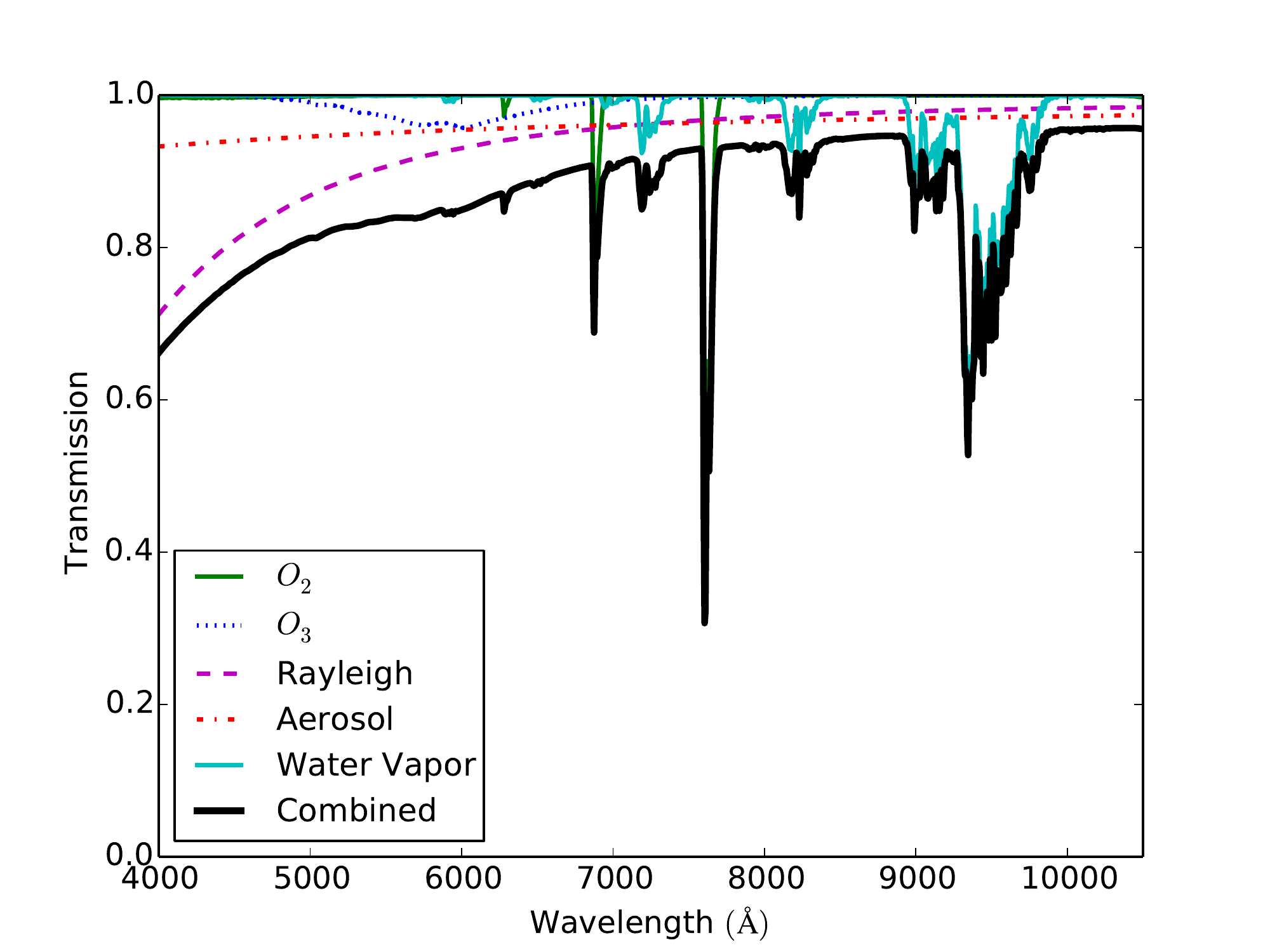}}
\caption{Standard atmosphere for the DES Y3A1 release computed with Gaussian 1nm FWHM smoothing.  The component values are
  listed in Table~\ref{table:STDatmo}.
\label{fig:standardAtm}}
\end{figure}

\begin{figure}
\scalebox{\figscalesmall}{\plotone{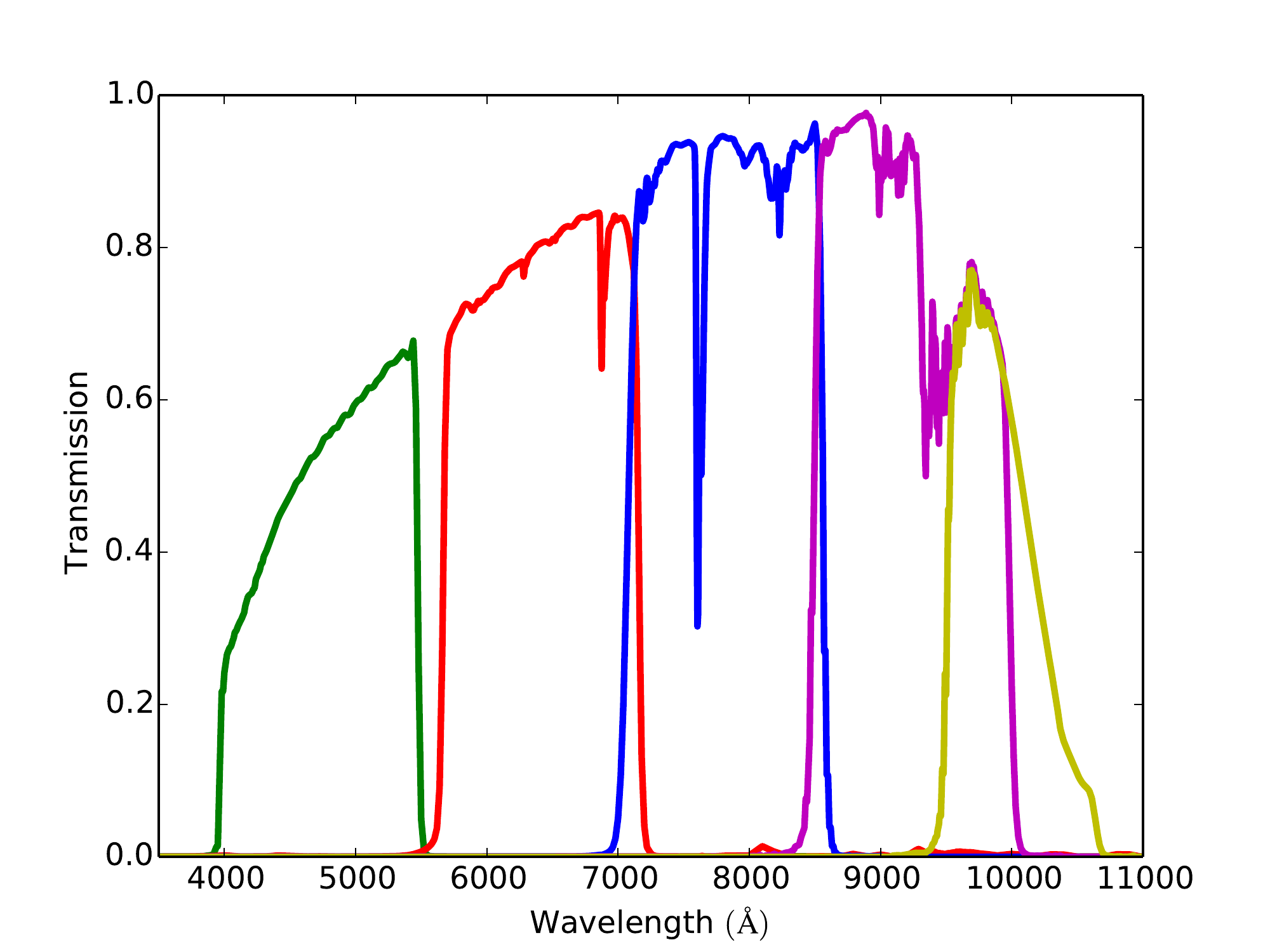}}
\caption{Standard passbands $\Sbstd$ for the DES Y3A1 release.  As with
  Figure~\ref{fig:Blanco/DECam_Passbands}, the lines are $g$-band (green), $r$-band (red), $i$-band
  (blue), $z$-band (magenta), and $Y$-band (yellow).}
          \label{fig:Standard_Passbands}
\end{figure}

Look-up tables of the $\Inaught^\obs$ and $\Ione^\obs$ integrals were computed at discrete points over a broad range of
the atmospheric parameter vector $\vec{P}^\atm$ using the focal-plane averaged instrumental passbands in Figure~\ref{fig:Blanco/DECam_Passbands}.
Variations across the focal plane in the $\Inaught^\obs$ values are corrected by
application of the superstar flats.
The $\Ione^\obs$ integrals are corrected for the variation of the wavelength profile of the instrumental passband 
across the focal plane by using data acquired with DECal for individual CCDs
and assuming the standard atmospheric parameters.
By definition, all of these corrections average to zero across the focal plane, so the standard passbands remain the reference.  
Interpolation of these discrete LUTs to continuous parameter space is done during the FGCM fitting and analysis procedures.

\section{Forward Global Calibration Process}
\label{sec:FGCMproc}

\subsection{Overview}
\label{sec:FGCMvue}

The FGCM includes several steps that are done once at the outset ({\it c.f.} Appendix \ref{app:init}).
This includes SQL queries of the ``FINALCUT'' catalogs produced by DESDM to
obtain both a ``demand'' list of exposures that are to be calibrated and a catalog of data from all observations of objects that are 
candidates to be used as calibration stars.
These observations are then cross-matched by location on the celestial sky to assign them to unique objects.
Selection is then made of candidate calibration stars, and a catalog of all observations of each candidate is created
that serves as the basis for the calibration fit and subsequent computation of calibration data products.

The FGCM does not use any prior knowledge of properties of potential calibration stars.
It begins with a bootstrap that uses the parameters of the standard passbands (Tables \ref{table:STDatmo} and \ref{table:STDparms})
as initial guesses for the model to be used to fit the observed data. 
Initial estimates of the $m_b^{\obs}$ magnitudes of the candidate calibration stars are made using a ``bright observation'' algorithm that
identifies groups of observations near the brightest observation found.
These are assumed to be approximately photometric,
and the algorithm yields estimates of the magnitudes of stars and rough estimates of residual ``gray'' errors on each exposure.

From this start, the process becomes cyclical with the steps illustrated in
Figure~\ref{fig:flowchart} (see also Appendix~\ref{app:build}):
\begin{enumerate}
\item Select calibration stars from those in the candidate pool with at least
  two observations in each of $griz$ (Fig.~\ref{fig:flowchart}(a)).
\item Select calibration exposures with at least 600 calibration stars from those in the campaign exposure demand list (Fig.~\ref{fig:flowchart}(b)).
\item Select ``calibratable'' nights with at least 10 calibration exposures from those in the campaign exposure demand list (Fig.~\ref{fig:flowchart}(c)).
\item Iteratively fit all $griz$ observations of all calibration stars on all calibration exposures taken on calibratable nights
to obtain best parameter vectors $\vec{P}^\atm$ and $\vec{P}^{\mathrm{inst}}$ (Fig.~\ref{fig:flowchart}(d)(e)(f)).
\item When the fit converges (or reaches a maximum number of iterations): compute best estimates for the magnitudes
  of calibration stars and dispersions of their repeated measurements, update
  estimates of residual ``gray'' extinction on individual exposures, and update the superstar flats (Fig.~\ref{fig:flowchart}(e)(g)(h)). 
\item If the sample of calibration exposures shows sign of residual ``gray'' loss of flux (see Sec.~\ref{sec:FGCMgrycor}),
then remove occulted exposures and start a new cycle with updated parameter vectors and analysis data products.
\end{enumerate}

\begin{figure*}
\plotone{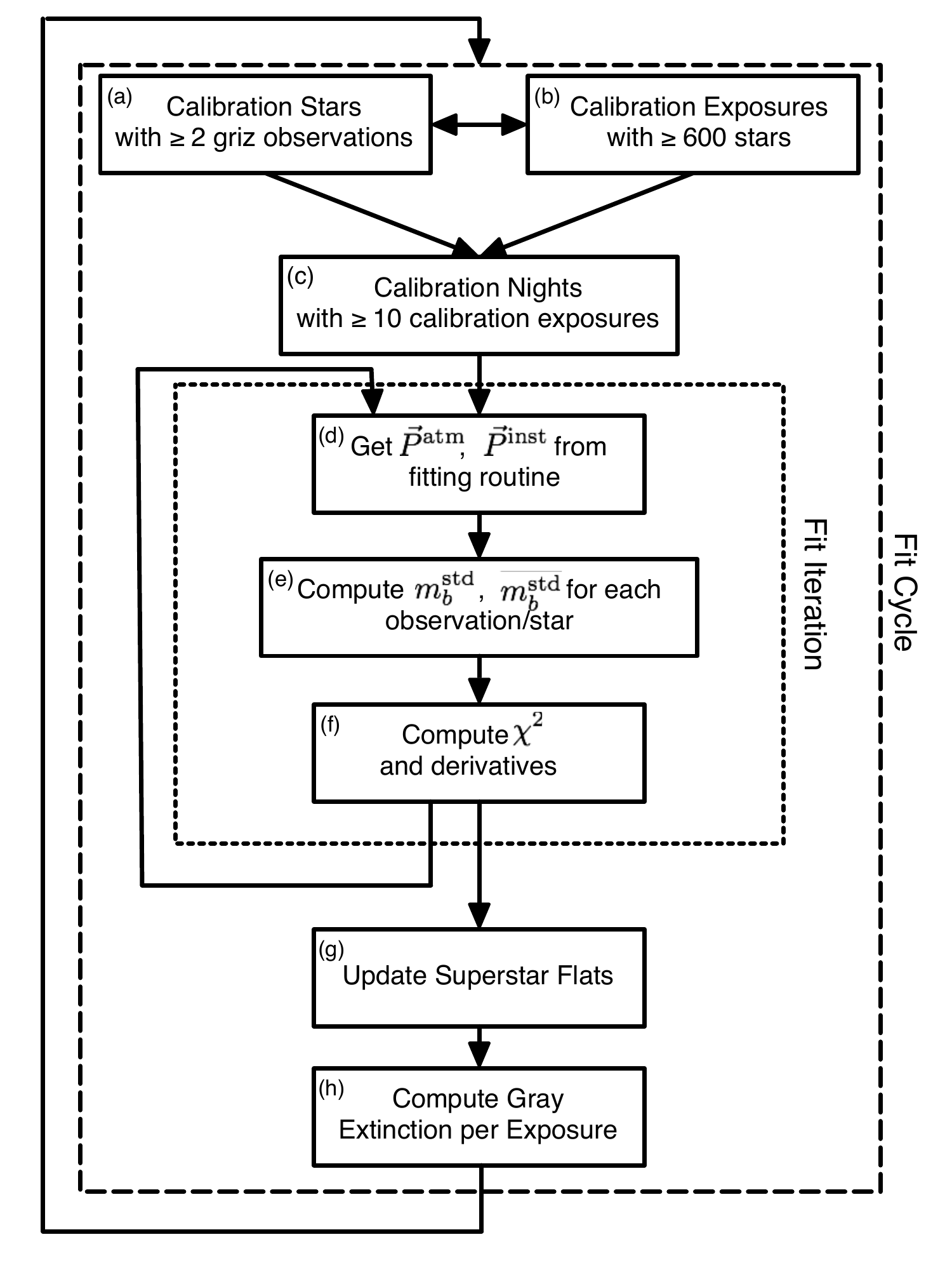}
\caption{Flowchart of FGCM selection and fit procedure.}
\label{fig:flowchart}
\end{figure*}

The $Y$-band observations are not used in the fit because those data are preferentially taken when observing is not optimal.
We ``dead reckon'' the $Y$-band magnitudes for calibration stars using the parameter vectors obtained from the fit to the $griz$ magnitudes:
\begin{enumerate}
\item Select $Y$-band calibration stars and exposures (Fig.~\ref{fig:flowchart}(a)(b)).
\item Compute $Y$-band magnitudes ($m_Y^\STD$) for the subset of $griz$ calibration stars that were also observed on $Y$-band calibration exposures (Fig.~\ref{fig:flowchart}(e)).
\item Compute $Y$-band dispersions, estimate ``gray'' extinction on individual exposures, and update $Y$-band superstar flats (Fig.~\ref{fig:flowchart}(g)(h)).
\item If the sample of $Y$-band exposures shows sign of significant loss of flux, then reselect $Y$-band calibration exposures and repeat. 
\end{enumerate}
We note that the precision of the final $Y$-band magnitudes is a useful internal ``blind''
diagnostic for the accuracy of the FGCM process (see Section~\ref{sec:precision}).

As a final step following the the fitting cycles
the FGCM process uses the final calibration star magnitudes $m_b^\STD$ and parameter vectors $\vec{P}^{inst}$ and
$\vec{P}^\atm$ to compute output data products for CCD images on science exposures in the campaign.
At this point the procedure:
\begin{enumerate}
\item Computes $\Inaught^\obs$ and $\Iten^\obs$ values from the FGCM fit parameter vectors.
\item Computes zero-point values and chromatic corrections from the FGCM fit parameter vectors ({\it c.f.} Eqn.~\ref{eqn:mstdfinal}).
\item Uses the standard magnitudes ($\overline{m^{\STD}}$) of calibration stars to estimate residual ``gray'' corrections.
\item Assigns quality flags and estimates errors of data products.
\end{enumerate}
These computations are described in more detail in Sec.~\ref{sec:ZPT}.

\subsection{The FGCM Fit}
\label{sec:FGCMfit}

The FGCM fitting step minimizes the weighted dispersion of repeated measurements of
the $m_b^{\STD}$ magnitudes ({\it c.f.} Eqn. \ref{eqn:mstdfinal}) of calibration stars
\begin{equation}
\label{eqn:chisq}
\chi^{2} = \sum_{(i,j)} {\frac{\left( m_b^\STD(i,j) - \overline{m_b^\STD(j)} \right)^2} {\sigma^\phot(i,j)^2}}, 
\end{equation} 
where the summation is over all calibration objects $j$ found on all $griz$ calibration exposures $i$.
The photometric error is defined as,
\begin{equation}
\label{eqn:sigphot}
\sigma^\phot(i,j)^2 \equiv \sigma^{\rm inst}(i,j)^2 + (\sigma_0^\phot)^2 , 
\end{equation}
where the instrumental error $\sigma^{\rm inst}(i,j)$ is computed by DESDM from source and background ADU counts, 
and the parameter $\sigma_0^\phot = 0.003$ is introduced to control possible underestimates of the errors assigned to the brightest objects.
This value is estimated from the residuals in the many exposures used to construct
the DESDM starflats as well as those taken in the supernova fields.
The error-weighted means of the calibrated magnitudes $m_b^\STD(i,j)$ of each calibration star $j$,
\begin{equation}
\label{eqn:mbar}
  \overline{m_b^\STD(j)} = \frac{\sum_{i} m_b^\STD(i,j)\sigma^\phot(i,j)^{-2}}{\sum_{i} \sigma^\phot(i,j)^{-2}},
\end{equation} 
are taken as the best estimates of the true standard magnitudes; here the summation is over exposures $i$ in band $b$.
The SciPy bounded fitting routine
FMIN\_L\_BFGS\_B\footnote{http://github.com/scipy/scipy/blob/v0.14.0/scipy/\allowbreak
  optimize/\allowbreak lbfgsb.py\#L47} \citep{zhu97}
is used to minimize the $\chi^{2}$ with the function value and derivatives with respect to all fit parameters explicitly computed.

The $\chi^{2}$ fitting statistic uses the standard magnitudes of stars with SEDs that span much of the stellar locus.
This provides sensitivity to both the amplitude and shape of the observing passband.
If the fit parameters are wrong for a given exposure, so too will be the chromatic corrections included in the computation of $m_b^\STD$.
As discussed in Appendix~\ref{app:chrocorr}, even within a single exposure there is typically sufficient range of
stellar spectra to constrain FGCM fit parameters with reasonable accuracy.
It is an important feature of the FGCM that it extracts as much information as possible
from each star that samples the observational passband of each exposure.

\subsection{FGCM Calibration Exposures, Calibratable Nights, and Gray Corrections}
\label{sec:FGCMgrycor}

As introduced in Sec.~{\ref{sec:clouds}, there are a number of factors that
affect the photometry that are not included in the FGCM fitting model.
A key to success of the FGCM fitting process is the ability to isolate a set of ``photometric'' exposures
free of clouds and significant instrumental errors.
To do this, the residual of each measurement $i$ of the magnitude of each calibration star $j$ is computed
using the parameter vectors from the most recent fit cycle,
\begin{equation}
\label{eqn:grayext}
E^\gray(i,j) \equiv  \overline{m_b^\STD(j)} - m_b^\STD(i,j).
\end{equation}
The average value of this residue is then computed for the calibration stars $j$ that are observed on
each CCD image of each candidate calibration exposure $i$,
\begin{equation}
\label{eqn:grayccd}
\CCD^\gray(i,{\rm ccd}) = \frac{\sum_{j} E^\gray(i,j)\sigma^\phot(i,j)^{-2}}{\sum_{j} \sigma^\phot(i,j)^{-2}},
\end{equation}
with statistical error,
\begin{equation}
\label{eqn:grayccdsigma}
\sigma^\phot(i,\CCD)^2 = \frac{1}{\sum_{j} \sigma^\phot(i,j)^{-2}}.
\end{equation}
The statistical error on $\CCD^\gray$ is typically $\sim 1-2\,\mathrm{mmag}$,
so structure on physical scales larger than the $\sim 0.2^{\circ}$ size of a DECam sensor can be resolved.
To take advantage of this, the average and variance of the residual extinctions of the CCD images on each exposure are computed as,
\begin{equation}
\label{eqn:grayexp}
\EXP^\gray(i) = \frac{\sum_\ccd \CCD^\gray(i,\ccd) \sigma^\phot(i,\ccd)^{-2}}{\sum_\ccd \sigma^\phot(i,\ccd)^{-2}},
\end{equation}
and
\begin{equation}
\label{eqn:grayvar}
\begin{split}
\VAR^\gray(i) &= \frac{\sum_\ccd \CCD^\gray(i,\ccd)^{2}
  \sigma^\phot(i,\ccd)^{-2}}{\sum_\ccd \sigma^\phot(i,\ccd)^{-2}} \\
& \qquad{} - \EXP^\gray(i)^{2}.
\end{split}
\end{equation}
Both $\EXP^\gray$ and $\VAR^\gray$ are used in the selection of calibration exposures to use in the FGCM fit ({\it c.f.} Appendix \ref{app:build}).
A night will be ``calibratable'' if a sufficient number of such calibration exposures were taken anytime during that night.
The $\CCD^\gray$ and $\EXP^\gray$ quantities are also used in the second step of the FGCM process to estimate the ``gray''
corrections for residual errors that are not included in the FGCM model (see Sec.~\ref{sec:fgcmgrycor}).

The FGCM yields detailed passbands for nearly all exposures taken on calibratable nights;
this includes exposures that were not used in the fit
as well as those that were.
These passbands can be used to compute both $\Inaught$ zero points for these exposures as well as
chromatic corrections either with known or hypothetical SEDs of sources (Eqn.~\ref{eqn:mstd}),
or linearized approximations based on the measured magnitudes of objects (Eqn.~\ref{eqn:mstdfinal}).

\begin{figure}
\scalebox{\figscalebig}{\plotone{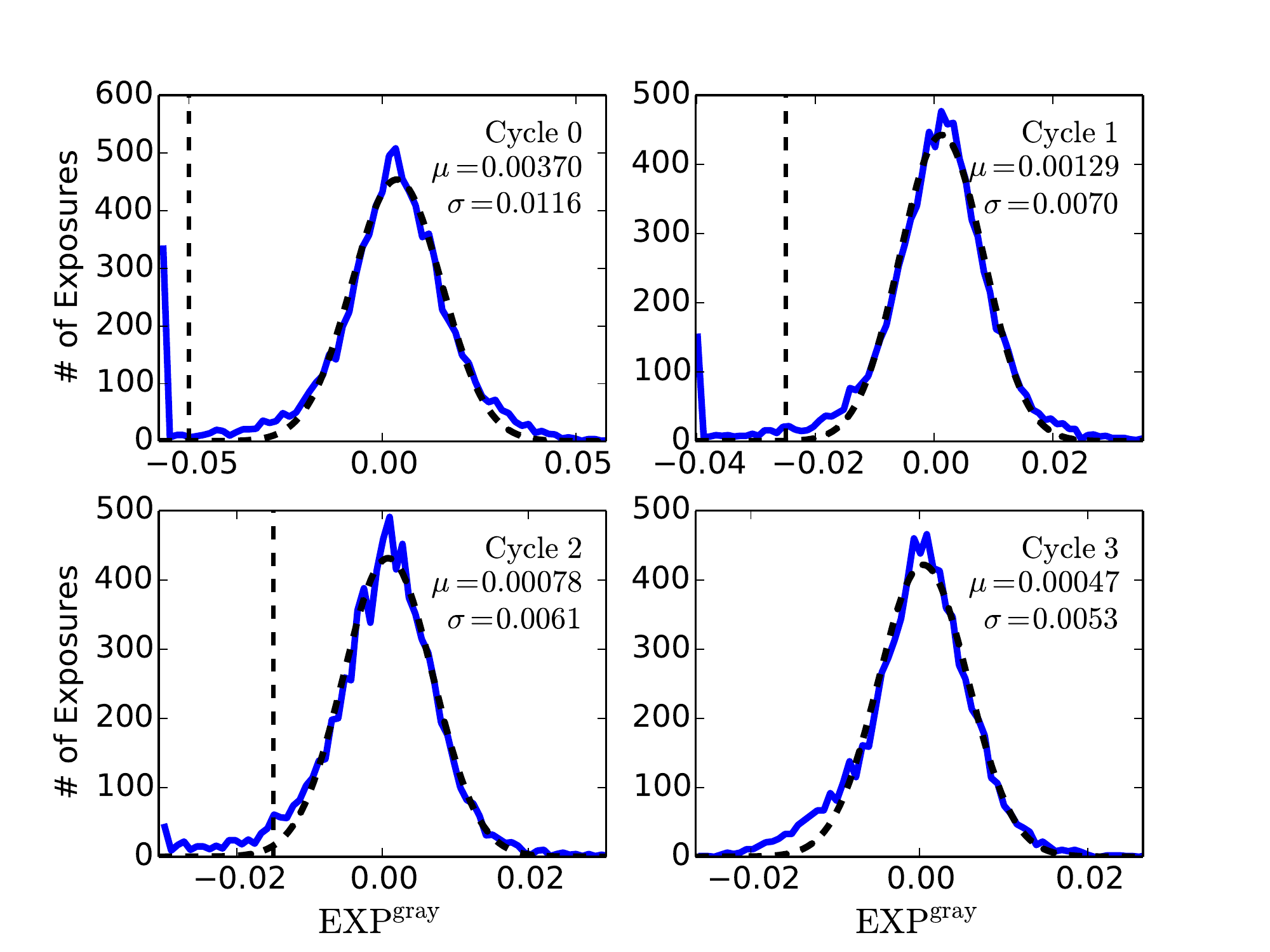}}
\caption{Extinction in candidate FGCM $i$-band calibration
  exposures. The observed distributions of $\EXP^{\gray}$ 
         (Eqn.~\ref{eqn:grayexp}) are shown in blue, while black dashed lines show Gaussian fits.
         Over or under flow counts are accumulated in bins at the extreme ends of the range.
         The top left plot is produced by the initial fit cycle that starts with default Standard passbands (Cycle 0).
         The asymmetric extended tail at negative values is due to cloud cover or instrumental error. 
         The subsequent plots show the progressive removal of exposures after continued fitting cycles sharpen the
         resolution and tighter cuts can be made to remove exposures with significant loss of flux.
         The cut values are shown as vertical dashed lines.
         The top right plot is produced by continued fit to the sample with exposures with $\EXP^\gray < -0.050$ removed after Cycle 0.
         The bottom left plot is produced after removing exposures with $\EXP^\gray < -0.025$ following Cycle 1. 
         The bottom right plot is the distribution produced by the final fit with exposures with $\EXP^\gray < -0.015$ removed after Cycle 2.}
\label{fig:FGCMgray}
\epsscale{1.0}
\end{figure}

\subsection{FGCM Y3A1 Fit Execution}

The Y3A1 fit was completed in four cycles.  
The distributions of the $\EXP^\gray$ obtained in $i$-band at the conclusion of these cycles are shown in Fig.~\ref{fig:FGCMgray}.
Similar distributions are obtained for all other bands; we show these for illustrative purpose.
There is clearly an asymmetric outlier population with significant loss of flux ($\EXP^\gray < 0)$ seen after the initial cycle
(top left panel in the figure).
These bias the computed magnitudes of the calibration stars, and prevent the fitting process from finding the optimal
solution for the passband parameter vectors.
Selection of calibration exposures at the end of each cycle of the fit is done by removing those found to be occulted
with a cut selected by examination of the non-occulted side of the distribution.
The precision of the subsequent cycle improves and allows the cut value to be tightened.
The FGCM fitting process is deemed to have converged when the distributions shown in the figure become nearly symmetrical,
and the bias of the distribution is reduced to an acceptable level.
We find that 2-3\% of candidate calibration exposures are removed on each cycle of the fitting process,
and that this is sufficient to reduce the bias by typically a factor of two.
After completion of the final fit Cycle 3, the bias of the fitted Gaussian peak has been reduced to $\mu \approx 0.5\,\mathrm{mmag}$,
and the width to $\sigma \approx 5\,\mathrm{mmag}$.

It can be seen in Figure~\ref{fig:FGCMgray} that the final sample includes a residual excess of exposures with some unaccounted loss of flux.
Cutting too tightly on $\EXP^\gray$ will bias the calibration star magnitudes in the sense opposite to that caused by exposures
taken through very thin cloud layers that remain in the calibration sample.
The FGCM minimizes this ambiguity by explicitly including in the final step of the process the ``gray'' correction introduced above. 

It is important to note that the magnitudes of the calibration stars ($\overline{m_b^{\STD}(j)}$) are not explicitly free parameters of the fit; 
they are functions of the observed ADU counts and the fitted parameter vectors $\vec{P}^{\mathrm{inst}}$ and $\vec{P}^\atm$.
So the $\chi^{2}$ function does not rigorously have the statistical properties of a chi-square,
but it is what we seek to minimize.
It is also very efficient and highly constrained. 
The FGCM calibration of the Y3A1 campaign used 2552 parameters
to fit 133,265,234 degrees of freedom (DOF),
and converged after four cycles (Fig.~\ref{fig:flowchart}) of 25, 50, 75, and 125 iterations.

\subsection{FGCM Y3A1 Fit Results}
\label{sec:y3a1results}

We provide here a summary of the statistics of the Y3A1 campaign and the parameters obtained from the FGCM fitting step.
A summary of statistics for the Y3A1 campaign is given in Table \ref{table:Y3A1stats}.
Query of the DESDM FINALCUT tables found demand for calibration of 41,562 $griz$ and 9770 $Y$-band wide-field and supernova field exposures
that were taken during the campaign\footnote{Standard star fields, taken at the
  beginning and end of each night, are not required for the science release and
  were not used in the calibration.}.
There were 11,710,194 candidate calibration stars found on exposures for which a calibration was requested.
At least one $griz$ calibration exposure was taken on 351 scheduled observing nights or half-nights,
while at least ten were taken on 335 nights. 
The Y3A1 FGCM calibration fit used 32368 $griz$ calibration exposures (78\% of the total) taken on 317 calibratable nights.
It produced standard standard magnitudes for 8,702,925 $griz$ calibration stars spaced nearly uniformly across the DES footprint 
($\approx$ 0.5 calibration stars per square arcminute).
Of these, 6,225,680 were also $Y$-band calibration sources.

\begin{deluxetable*}{lrl}
\tablewidth{0pt}
\tablecaption{DES Y3A1 Release Statistics}
\tablehead{
  \colhead{Statistic} &
  \colhead{Value} &
  \colhead{Comment}
}
\startdata
  Total exposures                        &   51332       &  DESDM demand file                \\
  $griz$ exposures                       &   41562       &                \\
  $griz$ cal exposures                   &   32368       &                \\
  $Y$ exposures                          &    9770       &                \\
  Nights with $>$ 1 cal exposure         & 351           &  Dome at least opened                    \\
  Nights with $>$ 10 cal exposures       & 335           &  Minimum to attempt calibration          \\
  Calibratable nights                    & 317           &  Some photometric time                    \\
  Number of $griz$ cal stars             & 8,702,925     &  Require $\ge$ 2 cal observations in each band       \\
  Number of $Y$ cal stars                & 6,225,680     &  A $griz$ cal star with $\ge$ 2 $Y$ cal observations  \\
  Number of $\ZPT^\FGCM$                    & 3,182,584     &  All CCD images (Table \ref{table:ZPTflags})    \\
\enddata
\label{table:Y3A1stats}
\end{deluxetable*}

\subsubsection{Superstar Flats}
\label{sec:results:superstar}

Five epochs of camera operations were identified and captured in the superstar flats $S_b^{\superstar}$.
These are not parameters of the fit, but are computed and updated after each calibration cycle from
the $\CCD^\gray$ values (Eqn. \ref{eqn:grayccd}) averaged over the DES wide-field calibration exposures taken in each band.
A typical $g$-band superstar flat, shown in Fig.~\ref{fig:superflatg}, is dominated by differences in the shorter wavelength sensitivity of the 
CCDs and their AR coatings, while the $i$-band superstar flat in Fig.~\ref{fig:superflati} exhibits smooth gradients of a percent or so across the focal plane.
These gradients are consistent with known ambiguities in the fitting technique
used to create the initial DESDM star flats.

\begin{figure}
\scalebox{\figscalesmall}{\plotone{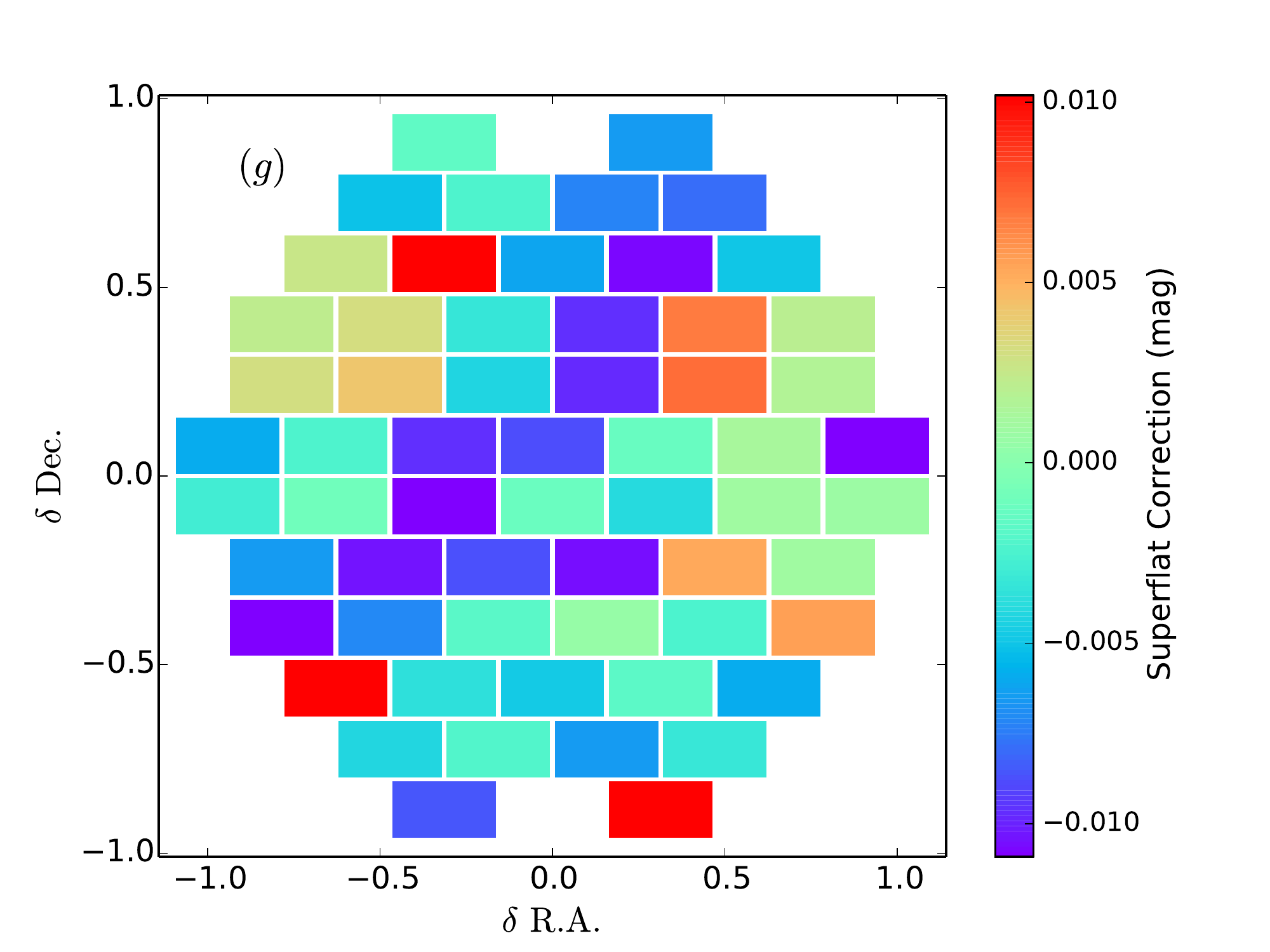}}
  \caption{Superstar flat in $g$-band derived from the third epoch of observing.  The axes are RA and DEC offset
        from the center of the field of view.
        The FGCM calibration process does not require these corrections to average to zero over the focal plane.}
  \label{fig:superflatg}
\end{figure}

\begin{figure}
\scalebox{\figscalesmall}{\plotone{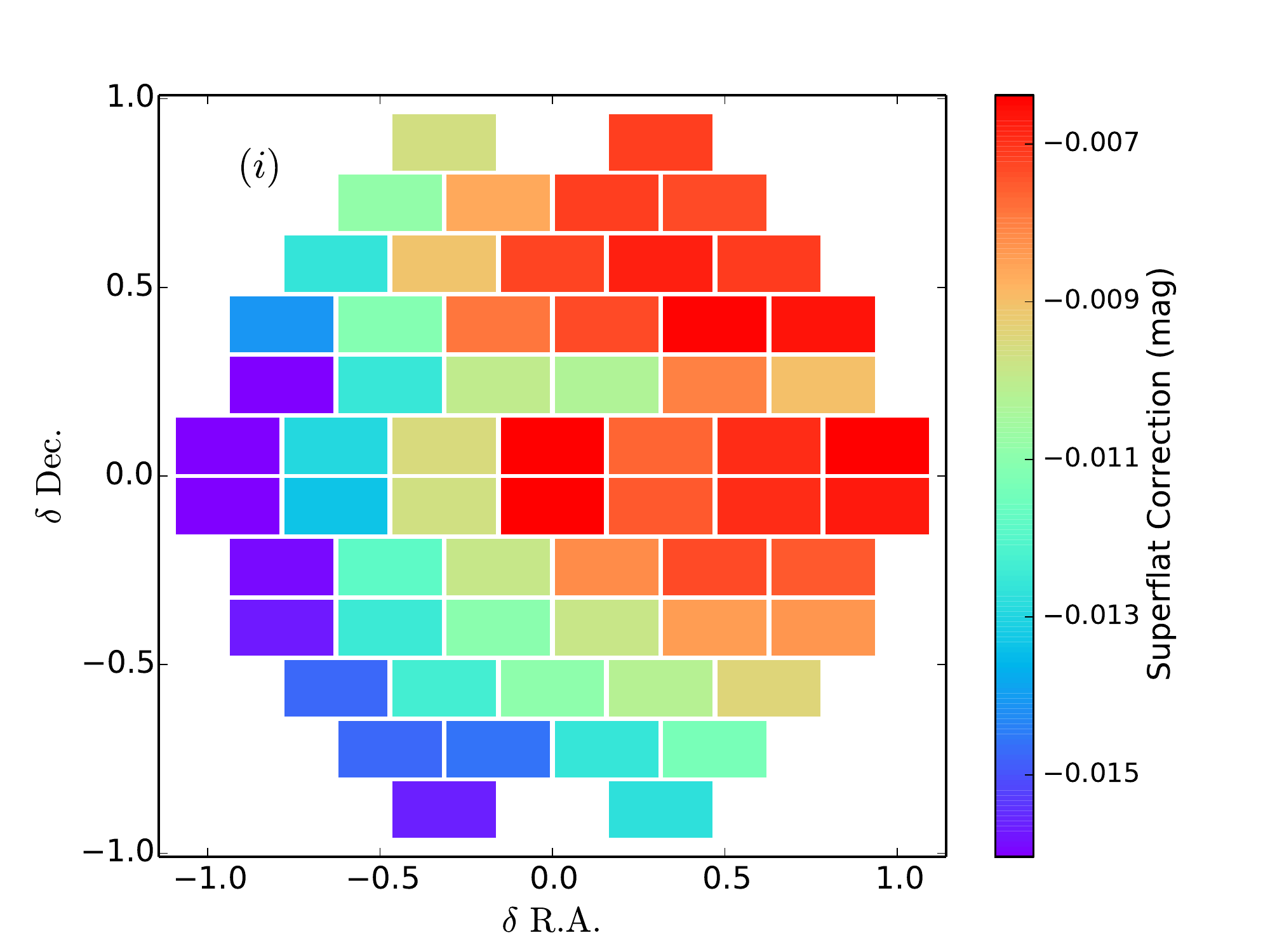}}
  \caption{Superstar flat in $i$-band derived from the third epoch of observing.}
  \label{fig:superflati}
\end{figure}

\subsubsection{Opacity Fit Parameters}
\label{sec:results:opacity}

The primary mirror was washed on seven dates during the three-year Y3A1 campaign, so the linear model used for
accumulation of dust $S^{\optics}$ requires 14 free parameters.
The resulting history, shown in Figure \ref{fig:FGCMoptics},
is consistent with laboratory engineering measurements taken on the wash dates.
It exhibits overall worsening of optimal transmission over time as expected for the aluminum mirror surface exposed to air,
and possible build up of dust on the downward-facing DECam external window that was not cleaned during this period of time.

\begin{figure}
\scalebox{\figscalesmall}{\plotone{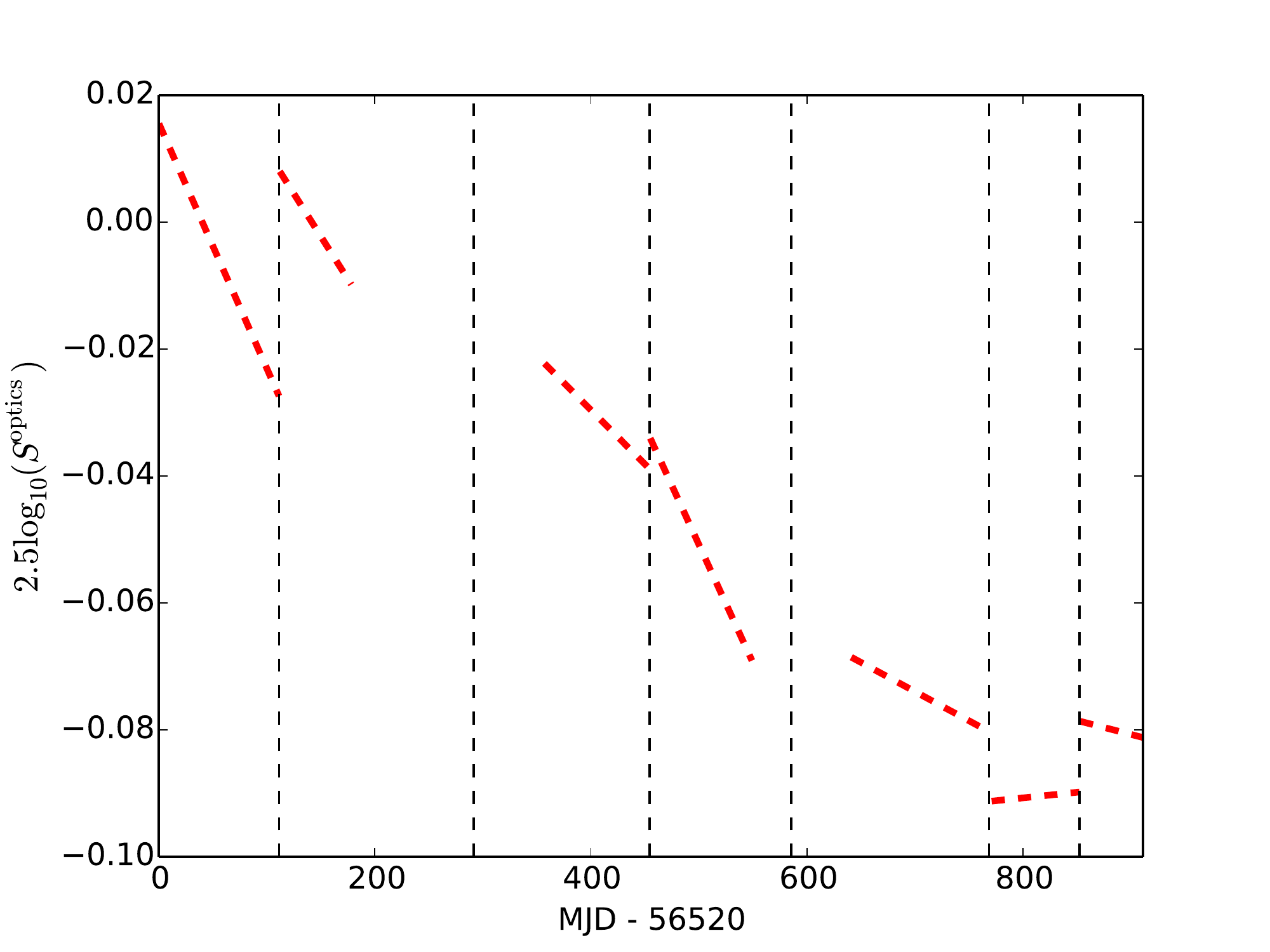}}
\caption{Throughput of the Blanco/DECam optical system.  The plot shows the piece-wise linear fit to the transmittance of the optics
        with discontinuities at the known dates when the primary mirror was
        washed (marked with vertical dashed lines and also once prior to the start of the survey that is not marked).
          \label{fig:FGCMoptics}}
\epsscale{1.0}
\end{figure}

\subsubsection{Atmospheric Fit Parameters}
\label{sec:results:atmosphere}

A summary of the atmospheric parameters obtained by the FGCM fit is given in Fig.~\ref{fig:FGCMatmo}.   
The auxiliary aTmCAM instrument was unavailable for the first year of the Y3A1 campaign,
and analysis of the data obtained in the latter two years was not available for inclusion in the Y3A1 calibration.
So these data are not used in the Y3A1 calibration. 
The SUOMINET GPS network provided measurements of atmospheric water vapor on 90\% of the nights of the DES Y3A1 campaign,
and these data were used to compute (Eq.~\ref{eqn:pwvaux}) most of the PWV values shown in the figure. 
The less precise linear form of Eq.~\ref{eqn:pwvut} was used for the remainder of the campaign observations. 
The aerosol depth values were fit in all cases with the linear form (Eq.~\ref{eqn:tauut}),
and as discussed in Sec.~\ref{sec:aerosol}, the aerosol optical index $\alpha$ is assumed to be constant during each night.
Note that for clarity the figure shows only nightly averages of the computed PWV and $\tau_{7750}$ values. 

A detailed analysis of the patterns and correlations in the meteorological parameters has not been done,
but the aerosol and water vapor distributions are consistent with historical data from the CTIO site.
There is evidence for seasonal variation in the fitted aerosol optical index consistent with smaller
sized particulates (larger optical indices) prevalent during the early spring start of the DES observing periods,
and particulates of larger cross sections dominate in the later summer periods.
Even with these trends removed the index remains noisy.
This may be due to the presence of multiple peaks in the likelihood function as would be the case if 
there were more than one component of aerosol particulate in the atmosphere (not an unreasonable expectation).
Future incorporation of data from the auxiliary aTmCAM instrument may allow inclusion of
two components of aerosol particulates in the atmospheric model.   
Residual error in these parameters remains reflected in the overall performance of the Y3A1 calibration that is discussed next.

\begin{figure}
\scalebox{\figscalesmall}{\plotone{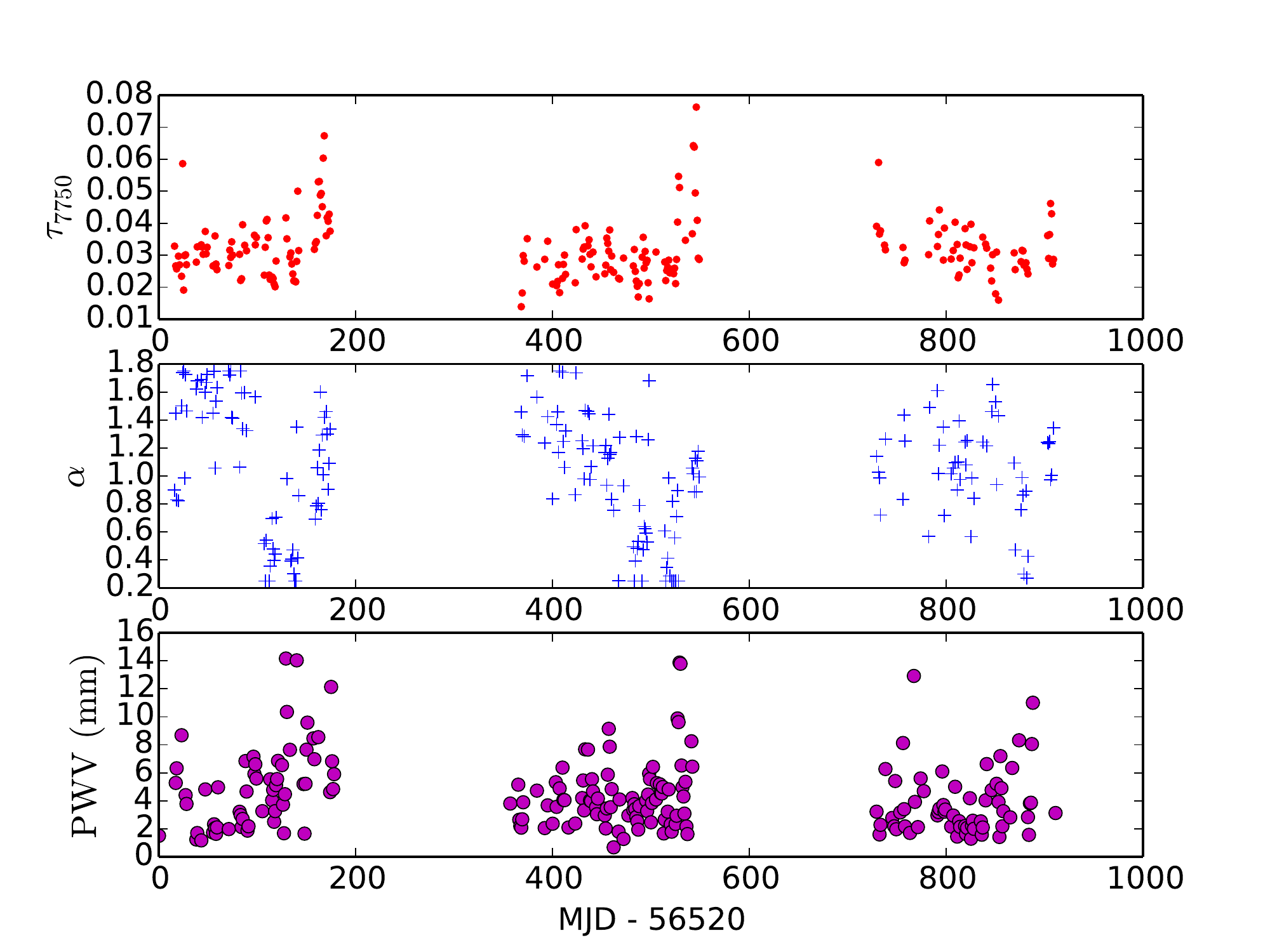}}
\caption{Atmospheric parameters from the Y3A1 FGCM fit.  The top plot shows the
  nightly average aerosol optical depth at $7750\,\mbox{\AA}$ for fitted
  exposures in the campaign.  The
  middle plot shows the aerosol $\alpha$.  The bottom plot shows the
  nightly average precipitable water vapor.  Only nights are shown
  where the given atmospheric parameter has an impact on the calibration, and
  therefore can be well fit.  For $\tau_{7750}$, only nights with at least 10
  exposures in each $g$ and $r$ are shown; for $\alpha$, only nights with at
  least 10 exposures in $g$ are shown; and for PWV, only nights with at least
  10 exposures in $z$ are shown.}
\label{fig:FGCMatmo}
\epsscale{1.0}
\end{figure}

\section{Performance of FGCM Calibration of Y3A1}
\label{sec:y3a1perf}

The metrics that we use to characterize the success of the FGCM fitting
procedure include the reproducibility of the calibrated $m_b^{\STD}$ magnitudes
of the calibration stars, and comparison with recently published Gaia $G$-band
data taken at the top of the atmosphere~\citep{gaia16}.
The first of these characterizes the ``precision'' or random error in
the nightly calibration vectors, and the comparison with Gaia is sensitive to
systematic errors in the fit that translate into uniformity variations.  We also evaluate the sufficiency of the linear
approximation to the chromatic correction for SEDs across the stellar locus.
We note that we also might use the uniformity of the observed stellar locus across the survey footprint
as a measure of the performance of the calibration~\citep[e.g.][]{ivezicetal04,highetal09,kellyetal14}.
Unfortunately the observed color distributions depend
on the resolution and accuracy of Galactic reddening corrections.
For the present work these are not sufficiently controlled at small spatial scales to address the sub-percent goals of the DES calibration.

\subsection{FGCM Fit Precision}
\label{sec:precision}

To evaluate the reproducibility (precision) of the FGCM calibration we consider the distributions of the
residuals $E^\gray(i,j)$ (Eqn.~\ref{eqn:grayext}) in $b = griz$ bands on exposures used in the final Cycle 3 of the Y3A1 FGCM fit.
Those measurements made with $\sigma^\phot < 0.010\,\mathrm{mag}$ are shown in Figure \ref{fig:FGCMdispersions}.
By design the fit minimizes the photon-statistics weighted variance of these
simultaneously in all bands. For diagnostic purposes we analyze these band-by-band   
\begin{equation}
\label{eqn:vargray}
\delta^2 E^\gray(b)   \equiv       \overline{ (E^\gray(b))^2 } - (\overline{ E^\gray(b) })^2.
\end{equation} 
We interpret these data in terms of the combined random errors in the FGCM fit without attribution to 
particular sources of these errors.

\begin{deluxetable*}{cccccc}
\tablewidth{0pt}
\tablecaption{Summary of FGCM Calibration Fit Results}
\tablehead{
  \colhead{Band} &
  \colhead{Mean Offset (mag)} &
  \colhead{Gaussian $\sigma$ (mag)} &
  \colhead{$\delta$FGCM (mag)} &
  \colhead{Fraction $<-2\sigma$} &
  \colhead{Fraction $>2\sigma$}
}
\startdata
g & $-0.00000$ & $0.0087$ & $0.0053$ & $0.036$ & $0.034$  \\
r & $-0.00000$ & $0.0080$ & $0.0045$ & $0.049$ & $0.051$  \\
i & $-0.00012$ & $0.0080$ & $0.0048$ & $0.045$ & $0.049$  \\
z & $0.00000$  & $0.0087$ & $0.0059$ & $0.034$ & $0.033$  \\
Y & $-0.00023$ & $0.0097$ & $0.0058$ & $0.020$ & $0.042$  \\
\enddata
\label{tab:dispersion}
\end{deluxetable*}

The DES survey is carried out in multiple ``tilings'' of the footprint,
and produces repeated observations of each calibration star that are generally well-separated in time;
a star is seldom observed in the same band more than once on a given night.
Exceptions are the supernova fields that are often observed with successive exposures when they are targeted.
With this exception, the errors in the FGCM fit parameters evaluated on different tilings are approximately independent of each other.
So we approximate the variances of the residuals as
\begin{equation}
\label{eqn:dispgray}
\delta^2 E^\gray(b)  \approx \delta^2 \FGCM(b) +  \frac{\sum_{(i,j)} \sigma^\phot(i,j)^2}{N_b},
\end{equation}    
where the sum is over the $N_b$ observations $i$ in band $b$ of all calibration stars $j$.
In this approximation $\delta^2 \FGCM(b)$ is the variance of the parent distribution of measurements of the magnitude of a star
introduced by random errors in the FGCM fit parameters.
We assume it is constant over the three-year survey and over the survey footprint.

After subtraction of the photon statistical uncertainties, 
values of $5-6\,\mathrm{mmag}$ are obtained for the precision $\delta \mathrm{FGCM}(b)$ of the Y3A1 fit in the $griz$ bands.
Table~\ref{tab:dispersion} summarizes the means and variances of Gaussian fits to the distributions,
and includes the fractions of observations found outside $2\sigma$ of the mean.
The Gaussian fits are reasonably good,
but the outlier populations are seen to be approximately twice that expected for purely random error.
These results are robust to variations in the cut on photon statistics over the range $0.005 < \sigma^\phot < 0.020$.
We note that these results are consistent with the precision implied by the residual exposure-averaged gray term shown in
Fig.~\ref{fig:FGCMgray} where the photon statistical errors are negligible. 
Analysis of the observations made of the DES supernova fields discussed in Sec.~\ref{sec:stability} below
support the hypothesis that the values determined from 
the entire data sample can be used to represent the error in any single measurement.

\begin{figure}
\scalebox{\figscalebig}{\plotone{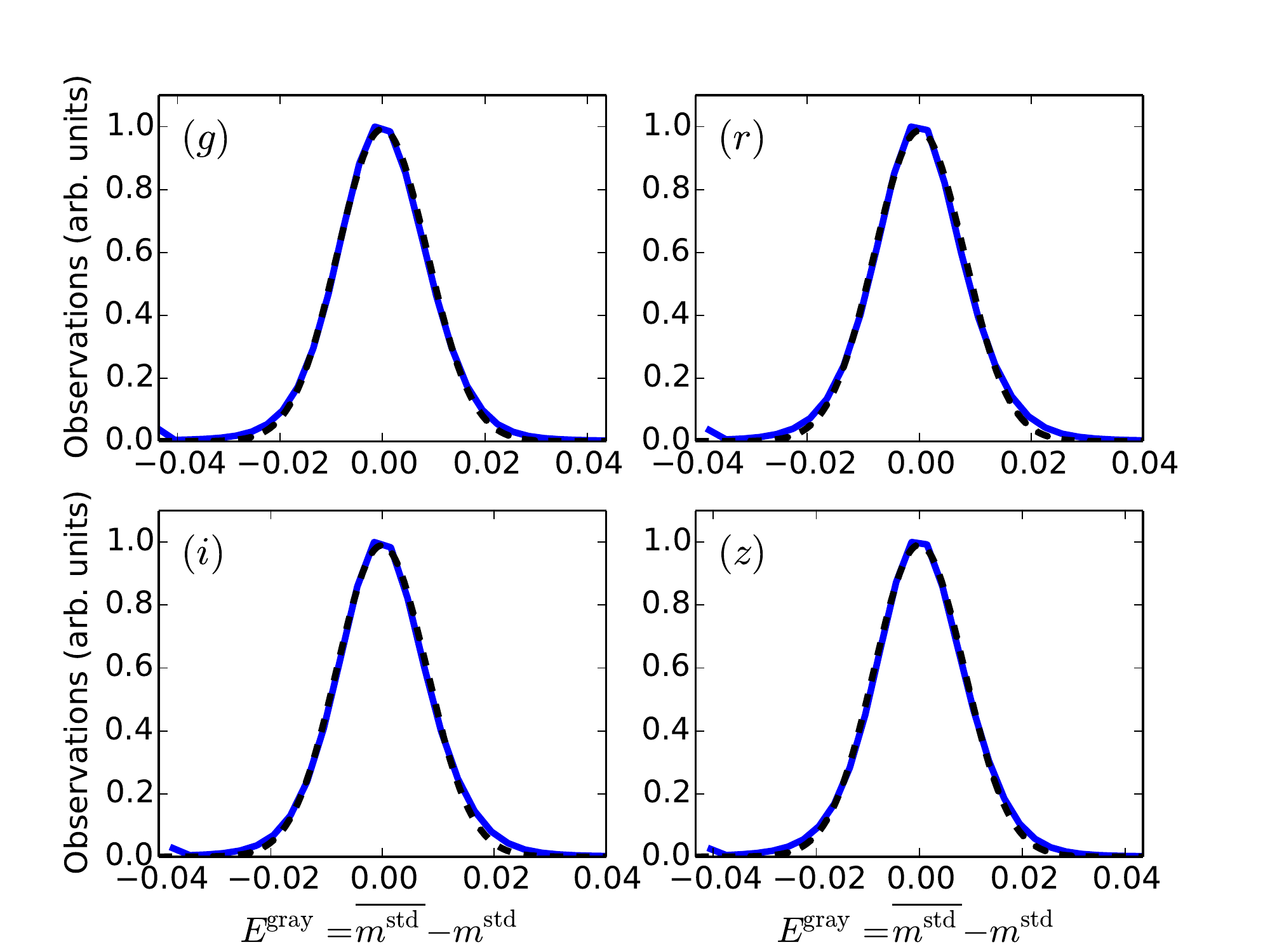}}
\caption{Dispersions of repeated measurements on calibration exposures of the $m_b^{\STD}$ magnitudes of
   calibration stars with $\sigma^\phot < 0.010$.
   The horizontal axis is $E^\gray(i,j)$ (Eqn.~\ref{eqn:grayext}), the difference between individual
   observations $i$ and the mean magnitude of all observations of the same star
   $j$.  The blue solid line is the histogram of the data, and the black dashed line is a
   Gaussian fit.  The calibration dispersion,
   $\delta\FGCM$ is computed by subtracting the estimated photometric errors in quadrature.
   Over or under flow counts are accumulated in bins at the extreme ends of the horizontal range.  The four plots correspond to the four 
   bands $griz$ used in the FGCM fit.  Each star is observed in nearly all
   cases only once on a given night, so the calibration errors of $5-6\,\mathrm{mmag}$ (Table~\ref{tab:dispersion}) 
   are indicative of the precision of the calibration of data taken on a single night.}
\label{fig:FGCMdispersions}
\epsscale{1.0}
\end{figure}

\subsubsection{The Y-Band Calibration}
\label{sec:Ybandprecision}

As described in Sec.~\ref{sec:FGCMvue}, the $Y$-band magnitudes are ``dead
reckoned'' from the atmospheric parameters derived from the $griz$ exposures.
Therefore, the $Y$-band data offer a useful internal check on the calibration
precision of the FGCM fit.  The subset of $griz$ calibration stars that are
also observed on at least two $Y$-band exposures are taken as candidates to be
$Y$-band standard stars.  Final $Y$-band calibration stars and exposures are
selected with a cyclical process to remove non-photometric exposures.  This
process is identical to that used for the $griz$ bands except that no
additional fits are made to the calibration parameter vectors.  The
$m_Y^{\STD}$ magnitudes are then computed from the $\vec{P}^\atm$ parameters
obtained from the $griz$ fit, and the distribution of residuals $E^\gray(i,j)$
are computed in the same manner as the $griz$ samples. The result is shown in
Figure~\ref{fig:Ydispersions} and included in Table~\ref{tab:dispersion}.  We
find that the $Y$-band calibration precision $\delta\FGCM(Y) =
5.8\,\mathrm{mmag}$ is comparable to that of the $griz$ bands.  This provides
assurance that the FGCM models and fitted parameters are sufficiently accurate
to account for sub-percent variations in the photometry in the reddest bands.

\begin{figure}
\scalebox{\figscalesmall}{\plotone{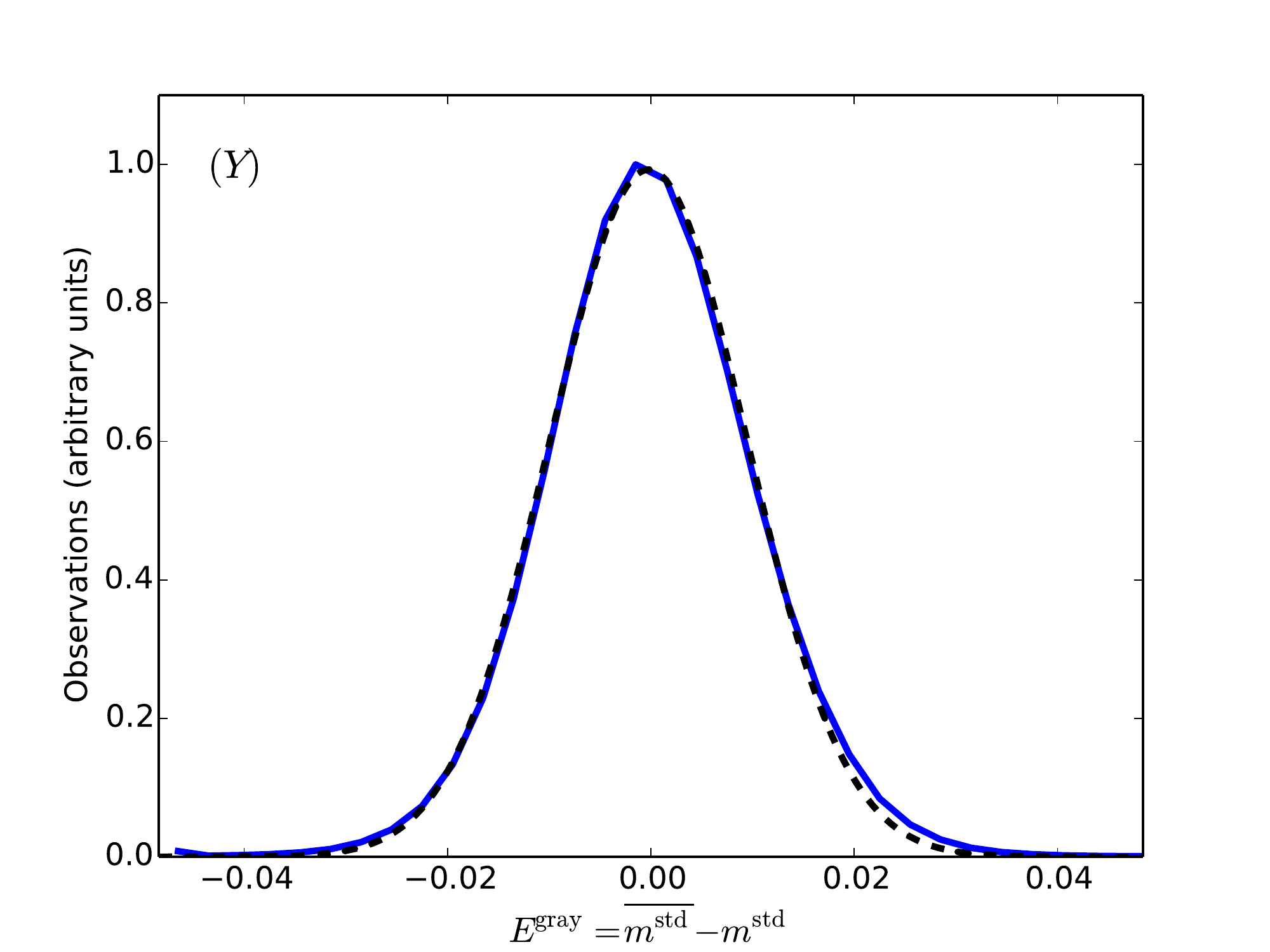}}
\caption{Dispersions of repeated measurements of the $m_Y^{\STD}$ magnitudes of $Y$-band
  calibration stars with $\sigma^\phot < 0.010$.
  The axes and data on the plots are the same as those of Fig.~\ref{fig:FGCMdispersions}.
  Subtraction of photon counting statistics in quadrature yields the value  $\delta \FGCM(Y) = 5.8$ mmag,
  in good agreement with the precision of the $griz$ data sets (Table~\ref{tab:dispersion}).}
\label{fig:Ydispersions}
\epsscale{1.0}
\end{figure}

\subsubsection{Precision of Calibration Star Magnitudes }
\label{sec:calstarprecision}

We examine in this section the internal precision with which the magnitudes of calibration stars are determined;
we note that, as in all cases, these magnitudes are only approximately normalized to an external scale.
With the assumption that the repeated measurements of the magnitudes of the calibration stars are independent of each other,
we estimate the random error in their mean magnitudes,    
\begin{equation}
\label{eqn:mbarerr}
\delta^2 \overline{m_b^\STD(j)} \approx \frac{\delta^2 \FGCM(b)}{({\rm N_b}(j) - 1)} + \frac{1}{\sum_i \sigma^\phot(i,j)^{-2}} +  (\sigma_0^{\STD})^2,
\end{equation}
where $N_b(j)$ is the number of observations $i$ of calibration star $j$ in band $b$. 
With the exception of the supernova fields, the number of tilings in the Y3A1 release is typically only 4-6.
So we reduce by one the number of effective samples of the fit error to account for the need to estimate 
the calibration star magnitudes from the averages of the measured values.
With this number of observations, the random error from the fit reduces to $< 4\,\mathrm{mmag}$ in all cases,
the statistical photometric error becomes $< 0.050\,\mathrm{mag}$ for nearly all calibration stars
(initially chosen with $\mathrm{S/N} > 10$ per exposure),
and for the brighter objects the overall errors approach the somewhat arbitrary control value $\sigma_0^{\STD} = 3\,\mathrm{mmag}$. 
The precision with which the calibration star magnitudes are known plays an important role in the
final assignment of calibration data products (Sec.~\ref{sec:fgcmgrycor}),
and in particular the ability to provide accurate calibrations of non-photometric exposures.

\subsection{FGCM Fit Stability}
\label{sec:stability}

The DES survey targets the supernova fields repeatedly every few days,
so these are used as a quality check on the stability of the FGCM calibration.
Shown in Fig.~\ref{fig:SNEoptics} are the $\EXP^\gray$ values (Eqn.~\ref{eqn:grayexp}) for all SNe calibration
exposures taken during the Y3A1 campaign.
Exposures in all bands are plotted in the figure.
The deviation of the mean of the residuals over the three-year survey is well below $5\,\mathrm{mmag}$.

\begin{figure}
\epsscale{1.0}
\scalebox{\figscalesmall}{\plotone{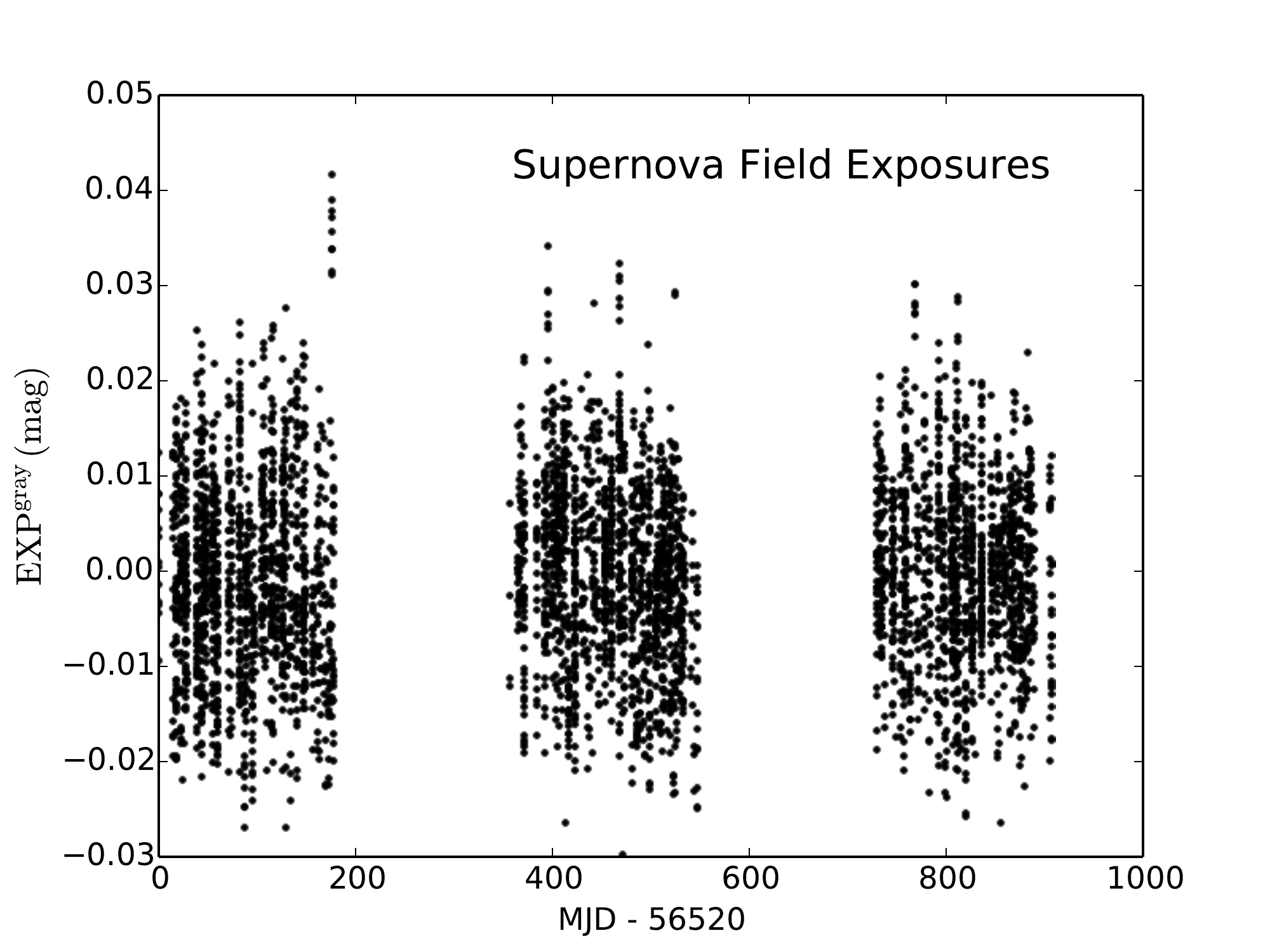}}
\caption{Average residuals of $griz$ magnitudes found on calibration exposures of the DES supernova fields over the full three Y3A1 campaign.
        These fields are observed at regular intervals during the campaign and demonstrate the stability of the FGCM calibration fit.
        Data from all four bands are included in the figure.}  
\label{fig:SNEoptics}
\epsscale{1.0}
\end{figure}

\subsection{FGCM Fit Uniformity}
\label{sec:uniformity}

The FGCM calibration will introduce correlations in the errors of measurements made closely spaced in time,
and these can be imprinted on the uniformity of the calibration error across the celestial sky by the survey tiling strategy.
The DECam focal plane is approximately $2^\circ$ in diameter, so it is expected that structure on this scale will appear
particularly in regions of the footprint that were not observed a large number of times in Y3A1. 

To look for possible spatial structure in the calibration we
compare with Gaia DR1 results \citep{gaia16} that were taken
above the atmosphere.  The Gaia $G$-band is a broad passband that is centered
approximately on the DES $r$-band and spans most of the DES $g$, $r$, and $i$ filters.
We fit a color transformation that combines weighted combinations of DES $g$, $r$, and
$i$ instrumental passbands to the Gaia $G$ published response.
Stars with DES color $0.5 < (g-i) < 1.5$ are spatially matched to stars in the Gaia catalog,
and the Gaia $G$-band magnitudes are compared with the transformed DES $gri$ magnitudes.
The mean differences of these magnitudes are binned
in HEALPIX pixels (NSIDE=256) and shown in Fig.~\ref{fig:gaia}.
The differences are
statistically well described by a Gaussian with $\sigma = 6.6\,\mathrm{mmag}$.  Spatial
structure at small and large scales can be seen at this level that is caused by
calibration and depth issues from both surveys.

\begin{figure*}
\scalebox{\figscalesmall}{\plotone{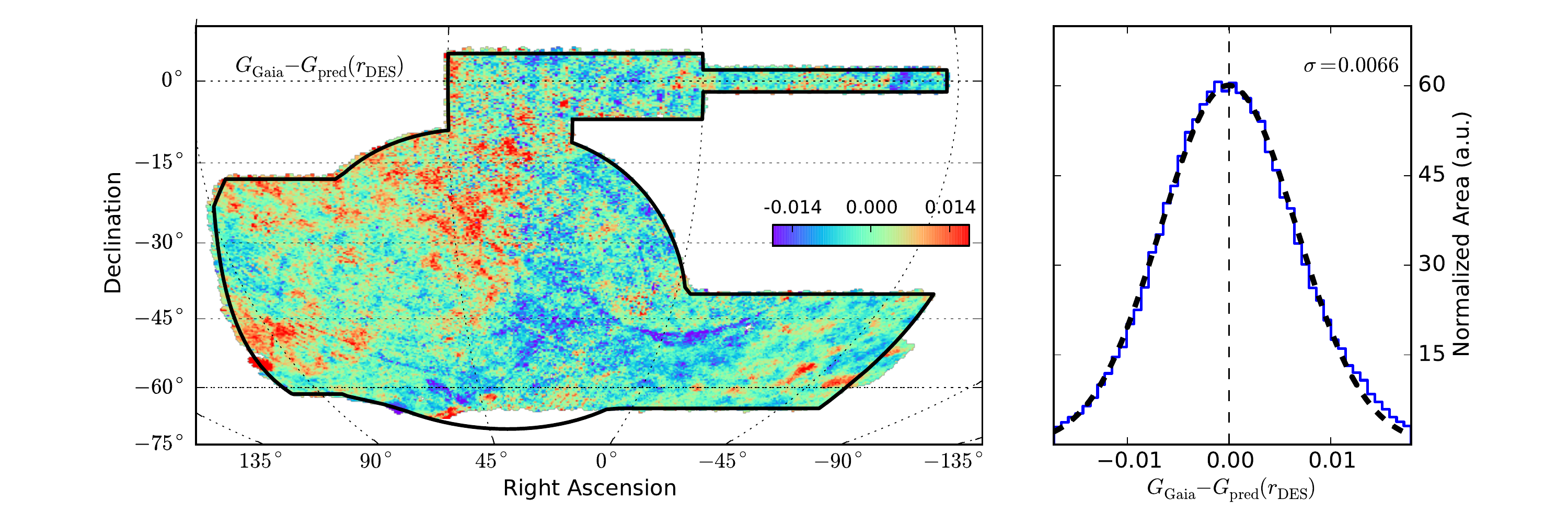}}
\caption{Offset map (left) and histogram (right) of difference between
  predicted Gaia $G$-band (from DES $r$-band) and observed Gaia $G$-band for
  stars with $0.5<(g-i)<1.5$, in pixels of HEALPIX NSIDE=256.
  Structure can be seen caused by calibration and depth issues from both surveys at small and large scales.
  A Gaussian fit with $\sigma = 6.6\,\mathrm{mmag}$ (black hashed curve)
  is shown on the histogram plot; overflow bins are not plotted.}
\label{fig:gaia}
\epsscale{1.0} 
\end{figure*}

\subsection{Linearized Chromatic Corrections}
\label{sec:chromcorrs}

The FGCM uses the linearized chromatic correction (Eqn.~\ref{eqn:mstdfinal}) 
to compute the standard magnitudes of calibration stars during the fitting stage.
Every evaluation of the $\chi^2$ requires these to be recomputed and
the fully integrated correction is computationally too slow to use for this purpose.
We discuss in this section both the size of the fully integrated correction and the accuracy of the linear approximation
for the stellar SEDs used in the fit.
Note that this does not address the accuracy of the calibration fit parameter vectors,
only the robustness of the linear computations.

To examine the accuracy of the linear approximation we use the stellar
spectral library of \citet{kellyetal14}, which combines SDSS spectra~\citep{sdssdr8} and the
Pickles spectral library~\citep{pickles98}.  For each template star, we
synthesize colors by integrating the SED with the standard FGCM passbands.  We
then randomly sample 50,000 exposure/CCD pairs from Y3A1 observing, and compute
the chromatic correction for each in two ways.  These are shown in
Figures~\ref{fig:chromapprox_blue}, \ref{fig:chromapprox_mid}, and
\ref{fig:chromapprox_red} for a sample blue star, middle-color star, and red
star, respectively.

First, we integrate the stellar spectrum with the passband for each exposure
and CCD combination.  These are plotted with red solid histograms in the
figures, and can be taken as estimates of the systematic chromatic offset that
needs to be included in the computation of the magnitude.
These corrections vary from as little as
$1-2\,\mathrm{mmag}$ ({\it e.g.} $i$ or $z$-band observations of blue stars) to as much as
$40-50\,\mathrm{mmag}$ ({\it e.g.} $g$-band observations of red stars).  Second, we use the
synthetic colors to estimate the linearized correction as done
during the FGCM calibration (Eqn.~\ref{eqn:mstdfinal}).  The residual of the linearized correction is
shown with the blue dashed histograms in the figures.  The median offset and
RMS estimated via median-absolute-deviation is also shown.

The linearized corrections reduce the residual error by factors of two to ten. 
In most cases the linearized correction reduces the systematic chromatic error to an RMS of $\sim2\,\mathrm{mmag}$,
though in some cases (particularly $g$-band for the reddest stars with $g-i \sim 3.0$),
the residuals have an rms of $\sim 5\,\mathrm{mmag}$.
These systematic uncertainties are reasonably well matched to the overall precision of the FGCM calibration.

\begin{figure}
\scalebox{\figscalesmall}{\plotone{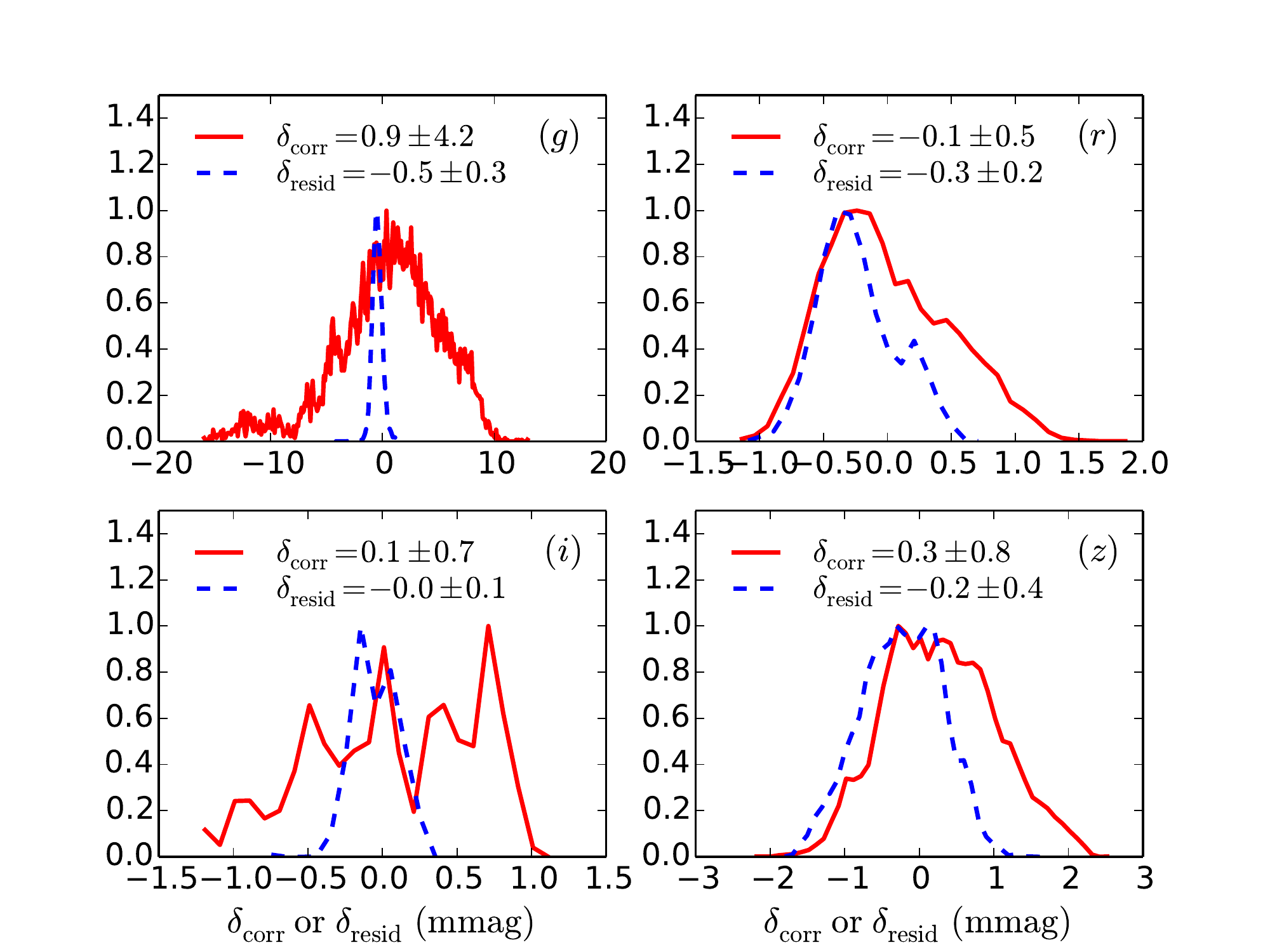}}
\caption{Chromatic corrections (red) and residuals for linearized corrections
  (dashed blue) for blue stars with $g-i \sim
  0.5$.\label{fig:chromapprox_blue}}
\end{figure}

\begin{figure}
\scalebox{\figscalesmall}{\plotone{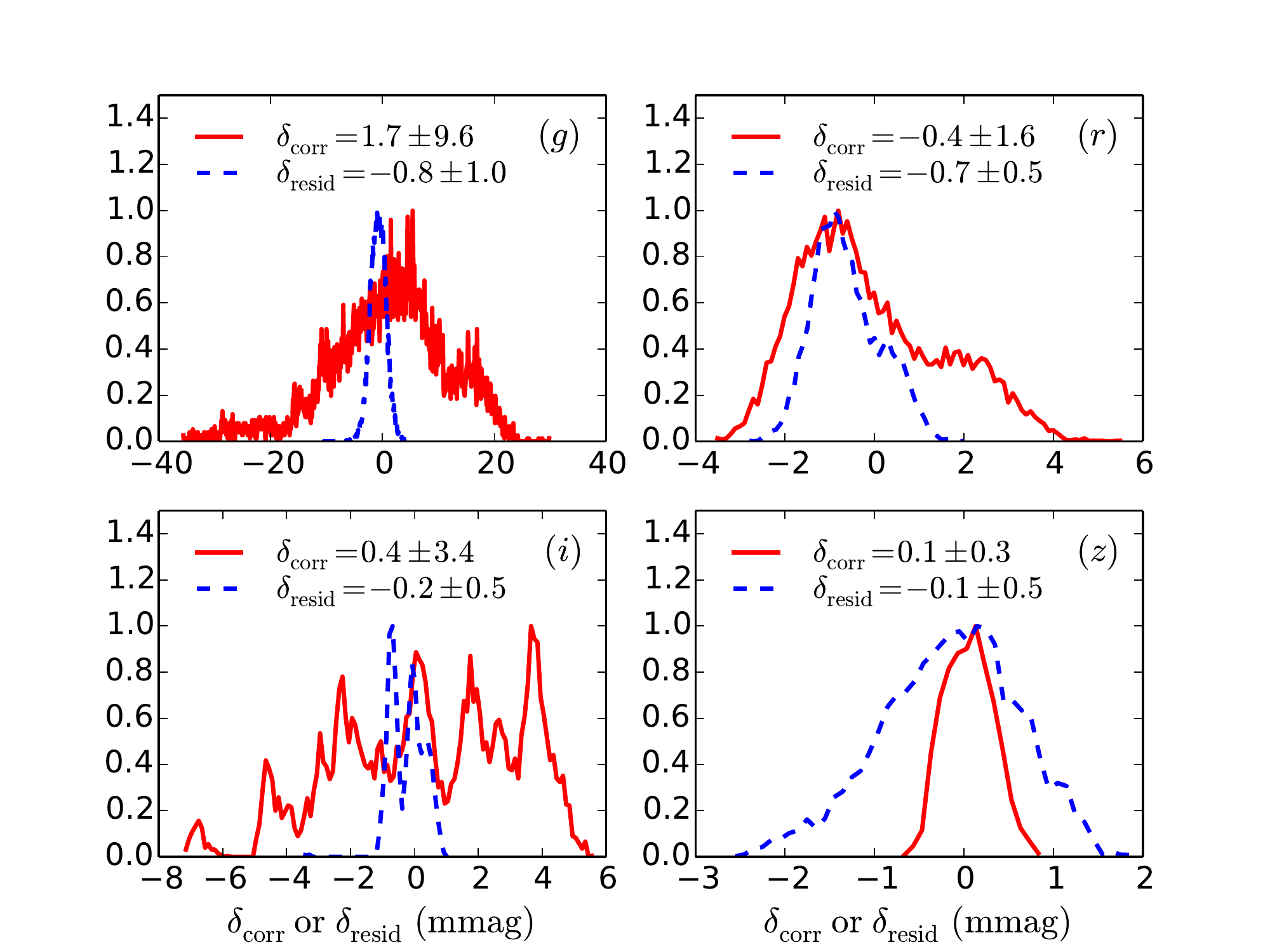}}
\caption{Chromatic corrections (red) and residuals for linearized corrections
  (dashed blue) for stars with $g-i \sim
  1.5$.\label{fig:chromapprox_mid}}
\end{figure}

\begin{figure}
\scalebox{\figscalesmall}{\plotone{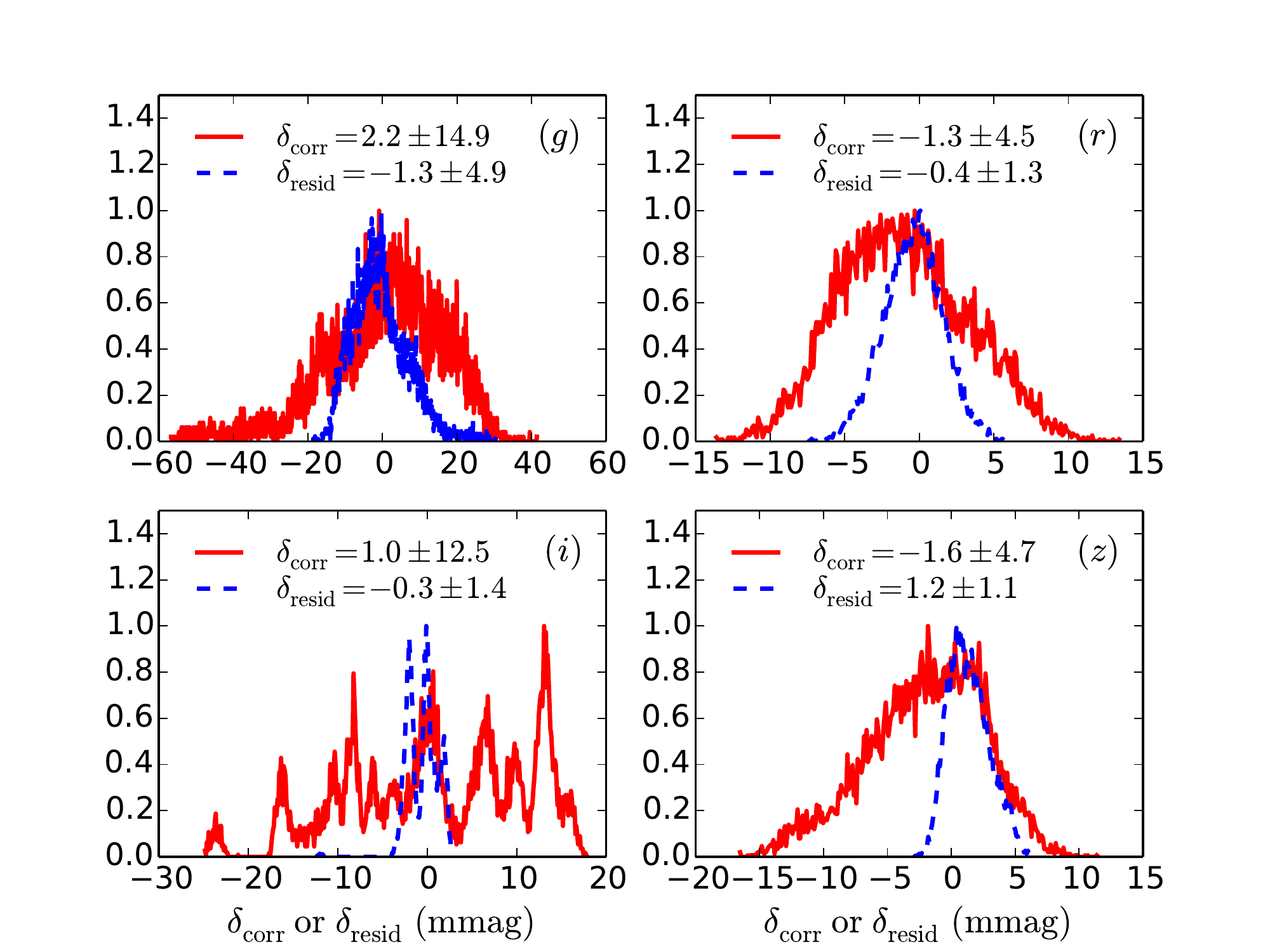}}
\caption{Chromatic corrections (red) and residuals for linearized corrections
  (dashed blue) for red stars with $g-i \sim
  3.0$.\label{fig:chromapprox_red}}
\end{figure}

While the previous analysis applies to single observations (as is the case with
transients such as supernovae), the impact of the linearized residuals on
coadd (average) magnitudes should be smaller if the corrections are
uncorrelated.  To test this, we start by computing the mean color of each
calibration star with no corrections.  After the $g-i$ color is computed, we match
this to the closest match in the \citet{kellyetal14} spectral library.
Chromatic corrections are then computed using the matched spectrum.
We then compute the offset between the coadd average
magnitudes using the linearized corrections and the integrated corrections for
blue, middle, and red stars as above.  Figure~\ref{fig:chromapprox_coadd_g_red}
shows a map of these offsets over the footprint (left panel), and a histogram of residuals
(right panel) for red stars in the $g$-band.
The linearized residuals for these stars are fit well by a Gaussian with $\sigma$ = $2\,\mathrm{mmag}$.
As can be seen in Fig.~\ref{fig:chromapprox_red} this case has the largest coadd residuals.
The residuals from the linearized corrections are $<0.7\,\mathrm{mmag}$ for every other band/color combination;
this validates our
assumption that use of the linearized correction in the fit does not produce significant loss of precision.

\begin{figure*}
\scalebox{\figscalesmall}{\plotone{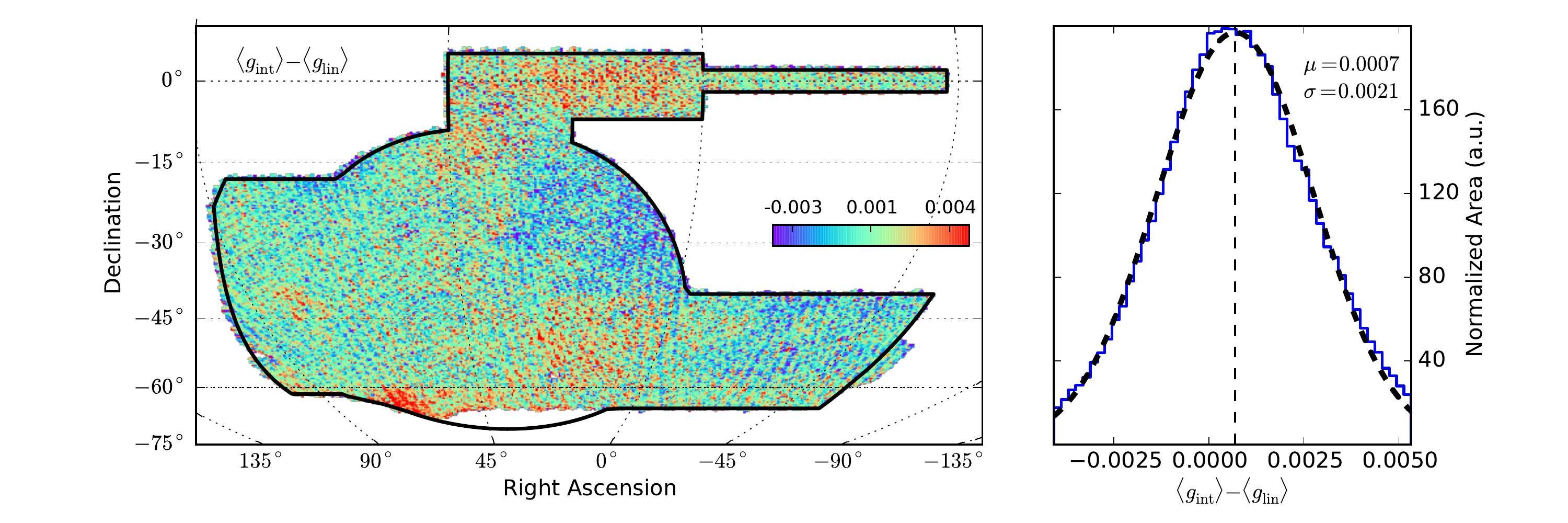}}
\caption{Offset map (left) and histogram (right) of the difference between the
  spectral-integration corrected coadd stellar magnitudes and the linearized
  approximation of the correction for red stars in the $g$-band.  While the Gaussian
  $\sigma$ of the residuals from the linearized corrections is $\sim2\,\mathrm{mmag}$ for this
  band/color combination, all other combinations have residuals of
  $<0.7\,\mathrm{mmag}$.\label{fig:chromapprox_coadd_g_red}}
\end{figure*}

\section{FGCM Data Products}
\label{sec:ZPT}

The FGCM fit yields a catalog of parameters $\vec{P}^\inst$
(Eqn.~\ref{eqn:FGCMinst}) and $\vec{P}^\atm$ (Eqn.~\ref{eqn:FGCMatm}) from
which it is possible to compute detailed passbands for observations taken on
calibratable nights.  It also yields a catalog of reference stars with
well-determined $grizY$ magnitudes in the FGCM standard system.  Both of these
catalogs are important products of the FGCM procedure that can be used in a
wide range of science analyses.

The FGCM also produces zeropoints to CCD images from Y3A1 exposures in the
initial query.  These zeropoints are defined to specifically calibrate object
magnitudes as measured in the DESDM FINALCUT processing.  These zeropoints
can be used to correct single-epoch images for studies of transient phenomena
or construction of multi-epoch coadd images and catalogs.

\subsection{FGCM Process Flags}

Assignment of values to FGCM data products follows one of several paths
identified by the parameter $\fflag$ summarized in Table~\ref{table:ZPTflags},
and detailed in this section.
Data acquired from $griz$ calibration exposures are given $\fflag = 1$, while
data from $Y$-band calibration exposures are given $\fflag = 2$.
Data from exposures that are not deemed to have been taken in photometric conditions, but that were taken on calibratable nights,
are in the $\fflag = 4$ category.  These first three categories comprise the exposures taken on calibratable nights that are themselves calibratable;
90.5\% of the exposures on the Y3A1 demand list are in one of these categories.
Objects observed on these exposures will have valid $m^\STD$ values with accurate chromatic correction.

There are a few individual CCD images that have a large number of
calibration stars, but were taken on nights with no calibration fit.  These are
$\fflag=8$ entries, and might be useful for some science cases, although the
chromatic corrections are purely instrumental.  A few percent of Y3A1
exposures that are in the initial demand list are found to be of too poor
photometric quality to complete a calibration; these are $\fflag = 16$ entries.
Finally, there are also a number of CCD images which were acquired on a
calibratable night, but contain too few calibration stars to determine reliable
``gray'' corrections.  These individual CCDs are retained as $\fflag = \fflag + 32$ for
completeness, but are not deemed science quality.

\begin{deluxetable*}{ccl}
\tablewidth{0pt}
\tablecaption{FGCM Calibration Quality Flags}
\tablehead{
  \colhead{$\fflag$} &
  \colhead{\% Y3A1 Exposures} &
  \colhead{Description}
}
\startdata
 1          &   77.7   &   CCD image on a $griz$ calibration exposure (\% of $griz$ exposures)      \\
 2          &   76.0   &   CCD image on a $Y$ calibration exposure (\% of $Y$ exposures)            \\
 4          &   13.2   &   CCD image taken on calibratable night, but not calibration exposure      \\
 8          &    0.4   &   CCD image recovered on a night with no calibration fit                   \\
 16         &    2.9   &   Unable to assign data products                                           \\
 33-36      &    6.1   &   CCD image on a calibratable night; unable to estimate gray correction    \\
 Any        &  100.0   &   All CCD images on all exposures                                          \\
\enddata
\label{table:ZPTflags}
\end{deluxetable*}

\subsection{FGCM Zero Points and Linear Chromatic Correction}

With sufficient knowledge of the SED of the target source, it is possible to
compute the full chromatic correction for observations taken on calibrated
nights ($\fflag = 1,\,2\,\mathrm{or}\,4$) using the fit parameter vectors.
However, the linear approximation to the chromatic correction can be easily
implemented, and is sufficient for many applications.
The FGCM provides calibration data products for both fully integrated and the linearized corrections
that are discussed in this section.

A calibrated measurement (on exposure number $\EXP$ and sensor $\CCD$) of the
$m_b^{\STD}$ magnitude of a celestial object can be computed as ({\it c.f.} Eq.~\ref{eqn:mstdfinal})
\begin{widetext}
\begin{equation}
m_b^\STD(\EXP,\CCD) = m_b^{\inst} + \ZPT^{\FGCM}(\EXP,\CCD) + 2.5 \log_{10}
\left( \frac{ 1 + \Fprime \times \Iten^\FGCM(\EXP,\CCD)} {1 + \Fprime \times \Iten^\STD(b) } \right),
\end{equation}
\end{widetext}
where $m_b^{\inst}$ is the instrumental magnitude computed by DESDM from
source and background ADU counts, nightly flats, and starflat corrections.
The zeropoint $\ZPT^{\FGCM}$ is computed from the integral of the
observing passband $\Inaught^{\obs}$ as in Eqn.~\ref{eqn:mtoa2},
and for reasons of flexibility includes the exposure time normalization
\begin{equation}
\label{eqn:zptfgcm}
\begin{split}
\ZPT^{\FGCM} &= 2.5 \log_{10}(\Delta T) + 2.5
\log_{10}(\Inaught^{\obs}(\EXP,\CCD)) \\
 & \qquad{} + \ZPT^{\gray} + \ZPTAB.
\end{split}
\end{equation}
\newpage

The passband integral $\Inaught^{\obs}(\EXP,\CCD)$ is computed as in Eqns.~\ref{eqn:optb} and \ref{eqn:Sinst}
\begin{equation}
\label{eqn:FGCMI0}
\begin{split}
\Inaught^{\obs}(\EXP,\CCD) & = S_b^\superstar(\CCD,\epoch) \\
& \qquad \times S^\optics(\vec{P}^\inst(\EXP), \mjd)  \\
  & \qquad \times \int_0^\infty S^{\mathrm{atm}}(\vec{P}^\atm(\EXP),\lambda)\\
& \qquad \times S_b^\DECal(\CCD,\lambda) \times \lambda^{-1} d\lambda,
\end{split}
\end{equation}
and the ``gray'' correction $\ZPT^{\gray}$ is described in the following subsection.

The normalized chromatic integral $\Iten^\FGCM(\EXP,\CCD)$ is similarly defined
\begin{widetext}
\begin{equation}
\label{eqn:FGCMI10}
\Iten^{\FGCM}(\EXP,\CCD) \equiv \frac{ \int_0^\infty
       S^{\mathrm{atm}}(\vec{P}^\atm(\EXP),\lambda) \times
         S_b^\DECal(\CCD,\lambda) \times (\lambda - \lambda_b) \lambda^{-1} d\lambda}
{\int_0^\infty
{S^{\mathrm{atm}}(\vec{P}^\atm(\EXP),\lambda) \times S_b^\DECal(\CCD,\lambda)
  \times \lambda^{-1} d\lambda}},
\end{equation}
\end{widetext}
with the cancellation of the superstar and optics terms that do not depend on
wavelength.  This integral is approximated via the LUTs described in Section~\ref{sec:stdsandluts}.

\subsection{FGCM Gray Corrections}
\label{sec:fgcmgrycor}

The FGCM takes advantage of the extensive network of calibration stars
with well-determined magnitudes to correct for residual errors due to effects
not included in the fit model.  With no guidance on possible wavelength dependence of
such failures, the algorithm uses the ``gray'' parameters computed after the
last cycle of the fitting process (Sec.~\ref{sec:FGCMfit}).  The
FGCM gray zeropoint correction, $\ZPT^{\gray}$, is determined from the stars on
the CCD whenever possible, an estimate from other CCDs on the same exposure will be used as a second choice,
and as a last resort $\fflag$ will be set to $\fflag + 32$.

For each CCD image, we compute $\CCD^{\gray}$ and $\sigma^{\phot}(\CCD)$ as
described in Eqns.~\ref{eqn:grayccd} and \ref{eqn:grayccdsigma}.  On CCD images
with at least 5 calibration stars and $\sigma^{\phot}(\CCD) <
5\,\mathrm{mmag}$, we assign the CCD estimate to the zeropoint gray value:
\begin{equation}
\label{eqn:fgcmgray}
\ZPT^\gray = \CCD^{\gray}(\EXP,\CCD).
\end{equation}
It is possible for a CCD image to contain few, if any, calibration stars even
though the entire exposure may contain CCD images with many stars (\emph{e.g.}
CCD images on exposures taken at the edges of the DES footprint).  In this
case, we compute $\EXP^{\gray}(\EXP)$ and $\VAR^{\gray}(\EXP)$ as in
Eqns.~\ref{eqn:grayexp} and \ref{eqn:grayvar}.  If there are at least 5 CCDs
with valid $\CCD^{\gray}$ values and variance $\VAR^{\gray} < 0.005^2$,
then we assign the exposure estimate to the zeropoint gray value:
\begin{equation}
\label{eqn:fgcmgrayexp}
\ZPT^\gray = \EXP^{\gray}(\EXP).
\end{equation}
No explicit limit is placed on the size or sign of the $\ZPT^{\gray}$ correction.
If both attempts fail, then the CCD image will be
uncorrected for gray extinction and the value of $\fflag$ will be increased by 32.

We note that application of the $\ZPT^{\gray}$ correction reduces to the
traditional use of a catalog of ``standard stars'' to estimate residual errors
in the calibration fit; in this case, the standard catalog is created by the
FGCM fitting step itself.  It relies on the presence of a sufficient number of
``photometric'' observations of these stars in the fitted data sample to
provide the needed reference magnitudes.  No chromatic correction can be
attached to a gray correction, but the retrieved chromatic integrals discussed in
Appendix~\ref{app:chrocorr} might usefully flag cases that have
significant residual chromatic effects; we have not yet studied this
possibility in detail.

\subsection{FGCM Calibration Errors}
\label{sec:FGCMerrors}

The FGCM assigns an error to the zeropoint $\ZPT^{\FGCM}$ (Eqn.~\ref{eqn:zptfgcm}) that includes
contributions from error in $\Inaught$ from the global calibration fit,
and from error in the gray correction $\VAR^{\ZPT}$ computed for each CCD image.
If $\ZPT^{\gray}$ is derived directly from $\CCD^{\gray}$, then the error in the gray correction 
is computed solely from photon statistics $\VAR^{\ZPT} = \sigma^{\phot}(\EXP,\CCD)^2$ (Eqn.~\ref{eqn:grayccdsigma}).
If it is derived from the average of other CCD images on the exposure, then $\VAR^{\ZPT} = \VAR^\gray(\EXP)$ (Eqn.~\ref{eqn:grayvar}).
Note that spatial structure in the residual errors on scales below the size of a CCD will
not be included in this estimate.

With the assumption previously discussed that the random errors in the FGCM fit
parameters from each tiling of the footprint are independent, the random error
in $\ZPT^{\FGCM}$ can be estimated,
\begin{equation}
\label{eqn:zpterr}
\begin{split}
\sigma^{\ZPT}(\EXP,\CCD) &= \bigg( \frac{\delta^2 \FGCM(b)}{N_{\mathrm{tile}}
  -1} \\
 & \qquad{} + \VAR^{\ZPT}(\EXP,\CCD) \\
 & \qquad{} + (\sigma_0^\ZPT)^2 \bigg)^{1/2},
\end{split}
\end{equation}
where the global fit error is $\delta\FGCM(b)$ (Eqn.~\ref{eqn:dispgray}), and
$N_{\mathrm{tile}}(\EXP,\CCD)$ is the average number of observations in band $b$ per
calibration star found on the CCD image (Eqn.~\ref{eqn:mbarerr}).  The
$\sigma_0^{\mathrm{ZPT}}$ systematic control term is again set to
$3\,\mathrm{mmag}$.  As discussed in Section~\ref{sec:FGCMfit}, exposures used
in the calibration fit (\emph{i.e.,} $\fflag <= 2$) were selected to have
little if any gray extinction, so the $\sigma^{\ZPT}$ values for these
exposures are typically $5\,\mathrm{mmag}$ or less.  While the $\ZPT^\gray$
corrections averaged over the focal plane can be equally accurate, errors in
magnitudes measured on non-photometric exposures (\emph{i.e.} $\fflag >= 4$)
can be significantly larger.

Shown in Figs.~\ref{fig:coadderrg} through \ref{fig:coadderrY} are the
calibration errors $\sigma^\ZPT$ averaged over exposures with $\fflag <= 4$ in HEALPIX pixels
across the full DES footprint that would be typical of the Y3A1 coadd catalog.
We show only the $g$, $i$, and $Y$ bands as examples, as the $r$ and $z$ bands
look very similar.
The structure seen in these plots is primarily due to correlations in the
varying number of tilings of regions on the sky.  For example, the supernova
fields can be readily identified as individual DECam focal plane footprints
with estimated errors near $3\,\mathrm{mmag}$ (the error floor set by $\sigma_0^{\mathrm{ZPT}}$) in $griz$-bands.

\begin{figure*}
\plotone{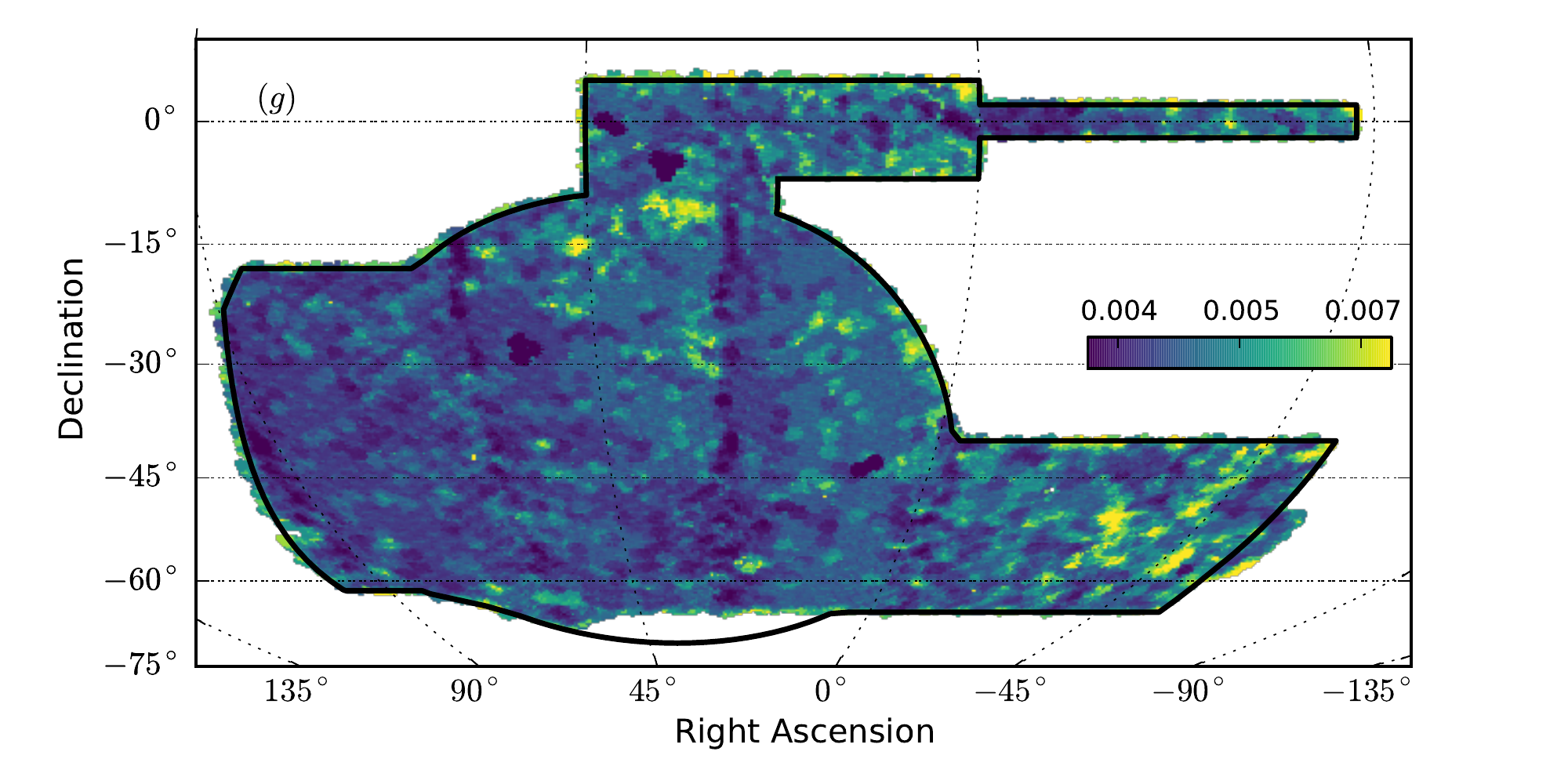}
\caption{Average calibration error $\sigma^\ZPT$ in $g$-band for catalog average
  stellar magnitudes found on exposures with $\fflag = 1$ or $\fflag = 4$, similar to
  the selection for the Y3A1 coadd catalog. The averages are binned at HEALPIX
  NSIDE=256.  The white region in the south was
  originally part of the DES footprint definition, but was eliminated to improve observing
  efficiency.}
\label{fig:coadderrg}
\end{figure*}

\begin{figure*}
\plotone{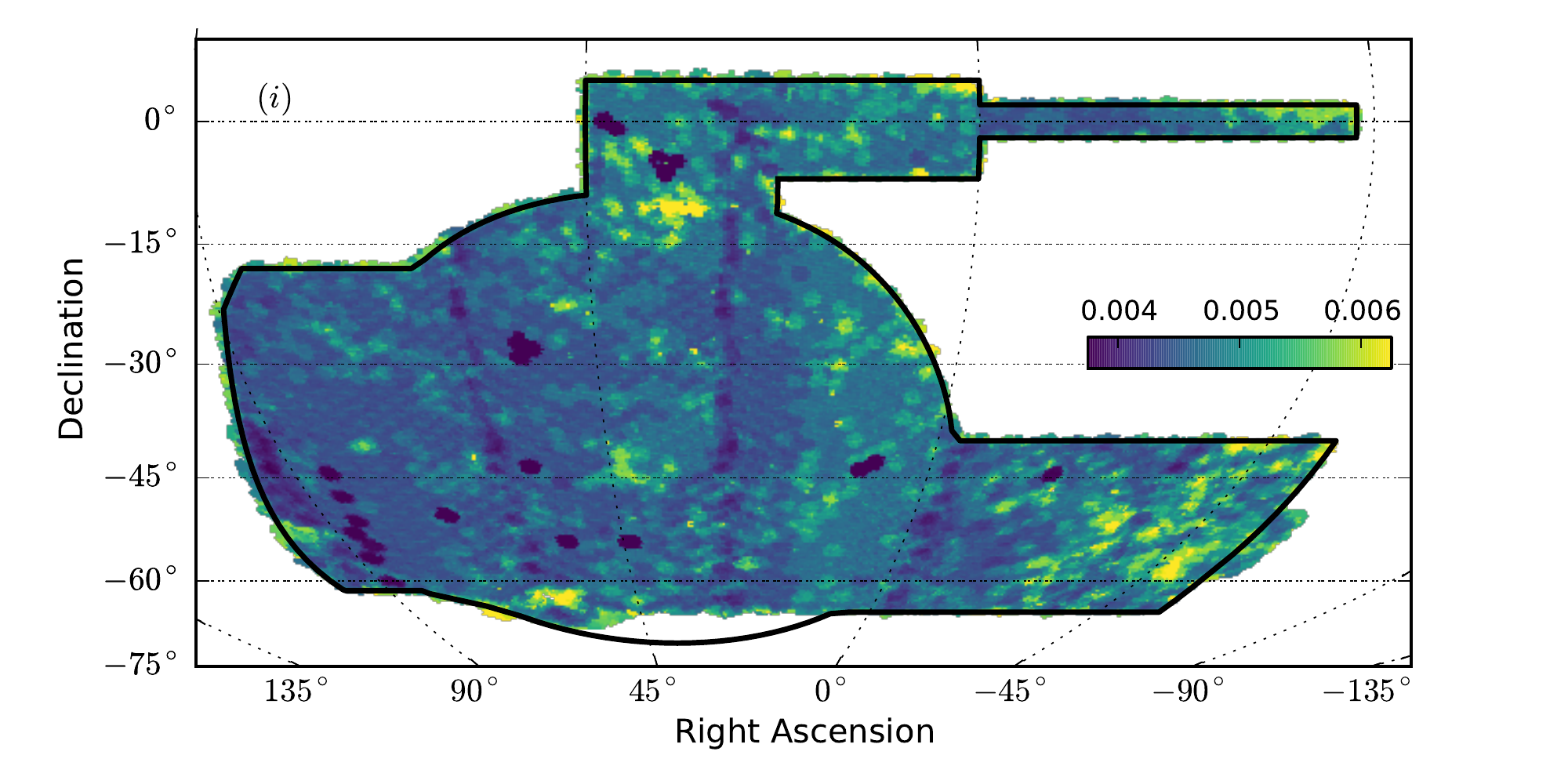}
\caption{Same as Figure~\ref{fig:coadderrg}, for $i$-band.\label{fig:coadderri}}
\end{figure*}

\begin{figure*}
\plotone{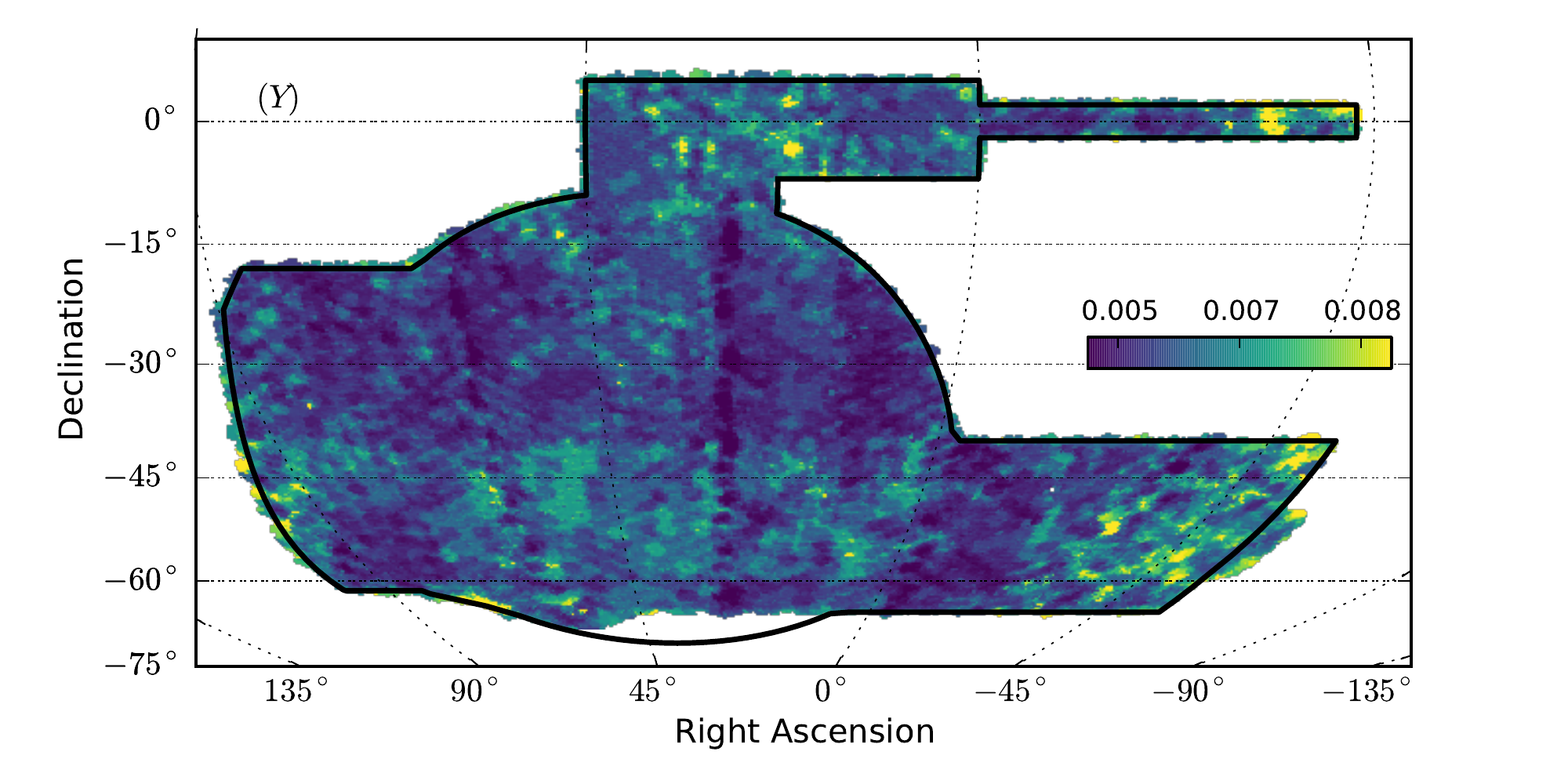}
\caption{Same as Figure~\ref{fig:coadderrg}, for $Y$-band.\label{fig:coadderrY}
  for exposures with $\fflag = 2$ or $\fflag = 4$.}
\end{figure*}

\section{Summary: Y3A1 and Beyond}
\label{sec:sumry}

We have presented a ``Forward Global Calibration Method (FGCM)'' for photometric calibration of wide-field surveys,
and we have presented results of its application to the first three years of the DES survey.
The FGCM combines data taken with auxiliary instrumentation at the observatory
with data from the broad-band survey imaging itself and models of the instrument and atmosphere
to estimate the spatial- and time-dependence of the passbands of individual DES survey exposures.
``Standard'' passbands have been chosen that are typical of the individual passbands encountered during the survey campaign.
The passband of any individual flux observation is combined with an estimate of the source spectral shape 
to yield magnitudes $m_b^{\STD}$ in the standard system. 
This ``chromatic correction'' to a standard system is necessary to achieve many DES scientific goals.

The FGCM achieves reproducible and stable photometric calibration of standard magnitudes $m_b^{\STD}$ of stellar sources
over the multi-year DES Y3A1 data sample with residual random calibration
errors $\sigma \approx 5-6\,\mathrm{mmag}$ per exposure (Table~\ref{tab:dispersion}).
The accuracy of the calibration is uniform across the 5000 square degree DES footprint
to within $\sigma = 7\,\mathrm{mmag}$ (Fig.~\ref{fig:gaia}).
The systematic uncertainty of magnitudes in the standard system due to the spectrum of the source
is less than $5\,\mathrm{mmag}$ for main sequence stellar spectra with $0.5 < g-i
< 3.0\,\mathrm{mag}$ (Sec.~\ref{sec:chromcorrs}).
A catalog of standard stars with well-known magnitudes in the DES standard system
is created by the FGCM procedure, as well as atmosphere models
for each exposure that allow computation of corrections of measured magnitudes to
the standard system.

Continued analysis of the calibration of the DES Y3A1 data set has pointed to several improvements that 
we anticipate installing in the FGCM in the future:  
\begin{enumerate}
\item We anticipate that incorporation of the aTmCAM data into the FGCM fit will yield
better reconstruction of the atmospheric conditions encountered during the survey.  
\item Selection of calibration objects at present does not attempt to remove variable sources (stars or quasars). 
Errors introduced into the calibration by such objects are presently included in our performance metrics.
While most DES images contain enough stars to realize some reduction in the average impact of these objects,
they may contribute especially to populations in the non-Gaussian tails of the residual distributions.
Known variable stars can be easily eliminated from the sample,
and relatively simple cuts based on light-curves observed by DES can be implemented to remove more of these objects.
\item The FGCM does not include any direct accounting of residual errors in the assignment of ADU counts to sources
due to errors in determination of the image point spread function (PSF) and variability of optical ``seeing''.
A first analysis of correlations between observed PSF values and residual photometric errors indicates these effects
are typically less than the $3\,\mathrm{mmag}$ control values used in the FGCM evaluations,
but in poor observing conditions can be worse.
Moreover, the DES strategy selects targets and filter bands based on observing conditions,
and so may introduce systematic bias in the calibration.
At present these errors are corrected only as part of the $\ZPT^\gray$ component in the $\ZPT^\FGCM$ values.   
A correction based on the PSF measured on each exposure and applied within the fitting cycle (e.g. as the superstar flats presently are)
can eliminate most of this effect and improve the convergence of the fit.
\item Transmission of out-of-band flux through the DECam optical filters is observed
in the $S_b^\DECal(\lambda)$ scans at the level of $\lesssim 0.1\%$.
This flux contributes to observed broad-band ADU counts.
The FGCM $\Inaught$ and $\Ione$ integrals are all computed over the wavelength interval from $3800\,\mbox{\AA}$ to $11000\,\mbox{\AA}$,
so they include this transmission to the extent that it is captured in the $S_b^\DECal(\lambda)$ data.
Full detailed DECal scans are time consuming, however,
and known errors in the existing scans introduce noise in the FGCM relative calibration.
While this noise is properly included in the FGCM Y3A1 performance metrics given in this report,
these errors will particularly affect chromatic corrections for non-stellar spectra,
and will complicate interpretation of absolute calibrations of the passbands.
We will acquire more accurate DECal scans over the out-of-band regions for analyses of these effects.
\item The FGCM uses the MODTRAN atmospheric transmission code to compute both the fit model
and corrections to observed broad-band magnitudes.
For the work here, computations were done with outputs smoothed with resolution of 1 nm (Gaussian FWHM).
Comparisons with other resolutions and codes will be done to determine sensitivity of the
calibration to the underlying computational methods for targets with SEDs of various types.
\item While not offering improved performance of the calibration, there is a simplification 
that can be made in the parameterization of the DES passbands. 
If we define $\lambda_b$ (Eqn.~\ref{eqn:lambdab}) with the standard passband (including the atmosphere)
rather than the instrumental passband, then $\Iten^\STD \equiv 0$ and would thus simplify Eqn.~\ref{eqn:mstdfinal}.
The magnitudes of the chromatic corrections are identical provided $\Iten^\obs$ and $\Iten^\STD$ are defined
consistently, so the final results of the calibration are unchanged, but the
formalism becomes more elegant.
\end{enumerate}

Finally we note that the concepts and techniques presented here will be even more powerful when applied to future data 
that will be obtained with the LSST.
The wide field of view and rapid cadence of the LSST survey will provide extremely fine and detailed sampling of observing conditions,
and the auxiliary instrumentation planned for the LSST observing site is designed to provide accurate determinations of 
changing passbands.
The LSST data set may be particularly well suited to implementation of some form of the exposure-by-exposure retrieval 
process discussed in Appendix~\ref{app:chrocorr} of this report.

\acknowledgments

The authors thank the staff of the Cerro Tololo Interamerican Observatory for
their expert and continuous support of the DES observing campaign. This work is
supported in part by the U.S. Department of Energy contract to SLAC
No. DE-AC02-76SF00515.

Funding for the DES Projects has been provided by the U.S. Department of Energy, the U.S. National Science Foundation, the Ministry of Science and Education of Spain, 
the Science and Technology Facilities Council of the United Kingdom, the Higher Education Funding Council for England, the National Center for Supercomputing 
Applications at the University of Illinois at Urbana-Champaign, the Kavli Institute of Cosmological Physics at the University of Chicago, 
the Center for Cosmology and Astro-Particle Physics at the Ohio State University,
the Mitchell Institute for Fundamental Physics and Astronomy at Texas A\&M University, Financiadora de Estudos e Projetos, 
Funda{\c c}{\~a}o Carlos Chagas Filho de Amparo {\`a} Pesquisa do Estado do Rio de Janeiro, Conselho Nacional de Desenvolvimento Cient{\'i}fico e Tecnol{\'o}gico and 
the Minist{\'e}rio da Ci{\^e}ncia, Tecnologia e Inova{\c c}{\~a}o, the Deutsche Forschungsgemeinschaft and the Collaborating Institutions in the Dark Energy Survey. 

The Collaborating Institutions are Argonne National Laboratory, the University of California at Santa Cruz,
the University of Cambridge, Centro de Investigaciones Energ{\'e}ticas, 
Medioambientales y Tecnol{\'o}gicas-Madrid, the University of Chicago, University College London, the DES-Brazil Consortium, the University of Edinburgh, 
the Eidgen{\"o}ssische Technische Hochschule (ETH) Z{\"u}rich, 
Fermi National Accelerator Laboratory, the University of Illinois at Urbana-Champaign, the Institut de Ci{\`e}ncies de l'Espai (IEEC/CSIC), 
the Institut de F{\'i}sica d'Altes Energies, Lawrence Berkeley National Laboratory, the Ludwig-Maximilians Universit{\"a}t M{\"u}nchen and
the associated Excellence Cluster Universe, the University of Michigan, the National Optical Astronomy Observatory, the University of Nottingham,
The Ohio State University, the University of Pennsylvania, the University of Portsmouth, 
SLAC National Accelerator Laboratory, Stanford University, the University of Sussex, Texas A\&M University, and the OzDES Membership Consortium.

The DES data management system is supported by the National Science Foundation under Grant Number AST-1138766.
The DES participants from Spanish institutions are partially supported by MINECO under grants AYA2012-39559,
ESP2013-48274, FPA2013-47986, and Centro de Excelencia Severo Ochoa SEV-2012-0234.
Research leading to these results has received funding from the European Research Council under
the European Union’s Seventh Framework Programme (FP7/2007-2013) including ERC grant agreements 240672, 291329, and 306478.

Facilities: \facility{BLANCO 4.0m/DECam}

\newpage
\appendix

\section{Estimation of SED Slope}
\label{app:SEDslope}
To complete the construction of a standard magnitude, we need a prescription for computation of the derivative of the SED of the source. 
We note that, if the passbands are flat or narrow in wavelength, then (though not rigorously) we can approximate, 
\begin{equation}
\label{eqn: Fprime}
\begin{split}
\Fprime \equiv \frac{F^{'}_\nu(\lambda)}{F_\nu(\lambda)} & = d \ln(F_\nu(\lambda)/d\lambda   \\
                                        & \approx {} -0.921 \frac{\Delta m^\obs}{\Delta\lambda}.
\end{split}
\end{equation}
We compute the slopes of the SED {\it at the boundaries between bands} as,
\begin{equation}
\label{SEDslope1}
\begin{aligned}
            S_0 &= -0.921 \times (m_r^\STD -  m_g^\STD) /(\lambda_r - \lambda_g)\\
            S_1 &= -0.921 \times (m_i^\STD -  m_r^\STD) /(\lambda_i - \lambda_r)\\
            S_2 &= -0.921 \times (m_z^\STD -  m_i^\STD) /(\lambda_z - \lambda_i)
\end{aligned}
\end{equation}
These lead to approximations for the slopes of the SED {\it across the passbands},
\begin{equation}
\label{SEDslope2}
\begin{aligned}
            \Fprimeraw(\lambda_g) & \approx S_0 - 1.00 \times ( (\lambda_r - \lambda_g)/(\lambda_i - \lambda_g) ) \times ( S_1 - S_0 )\\
            \Fprimeraw(\lambda_r) & \approx (S_0 + S_1)/2.0\\
            \Fprimeraw(\lambda_i) & \approx (S_1 + S_2)/2.0\\
            \Fprimeraw(\lambda_z) & \approx S_2 + 0.50 \times ( (\lambda_z - \lambda_i)/(\lambda_z - \lambda_r) ) \times ( S_2 - S_1 )\\
            \Fprimeraw(\lambda_Y) & \approx S_2 + 1.00 \times ( (\lambda_z - \lambda_i)/(\lambda_z - \lambda_r) ) \times ( S_2 - S_1 ).   
\end{aligned}
\end{equation}
The ``fudge factors'' (-1.00, 0.50, and 1.00) are used for the bands at the end of the spectrum
to accommodate extrapolation across passbands that are not flat nor narrow. 
The empirical determination of these factors and the accuracy of Eqns.~\ref{SEDslope2} are discussed in Appendix~\ref{app:chrocorr}.

\section{FGCM Chromatic Corrections and Retrieval}
\label{app:chrocorr}

We discuss in this appendix a method to retrieve the chromatic integrals $\Inaught$ and $\Ione$ for individual exposures given
a sufficiently large set of well-calibrated stars.
This method might be used retroactively to improve the temporal frequency of a calibration done initially on 
a nightly basis such as has been done for Y3A1.
It is possible to extract the value of the two passband integrals for each individual exposure from 
the behavior of the observed flux produced by stars of different colors.
This is highlighted in Figure \ref{fig:starchrom} which shows the dependence of the chromatic corrections made in the
Y3A1 calibration for stars of differing colors on two $z$-band exposures, with high and low PWV values.

\begin{figure}
\plotone{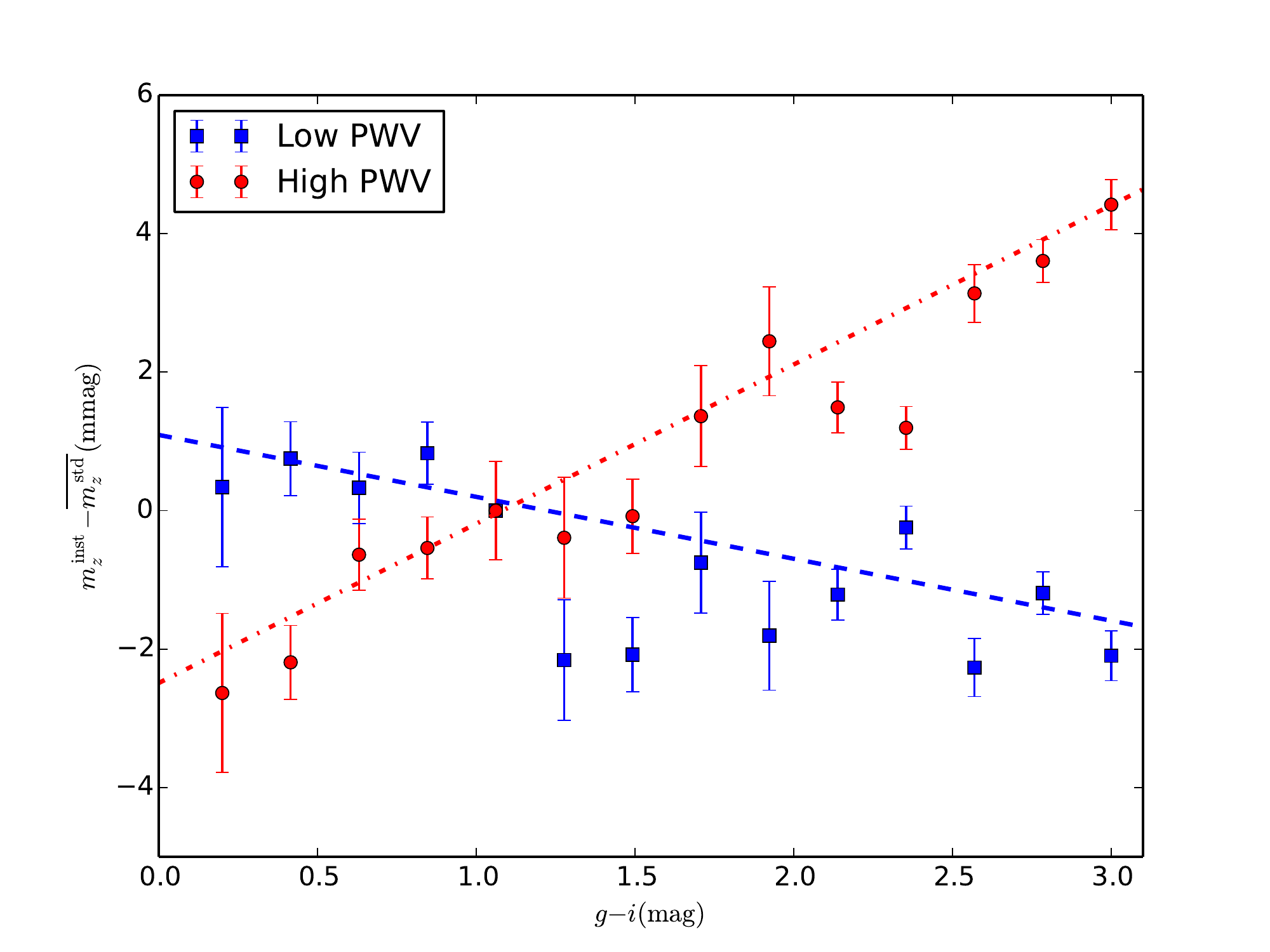}
\caption{Dependence of the chromatic corrections for stars of differing
  colors.  In this example, two $z$-band exposures were chosen, one with high $\PWV$
  (red circles), and one with low $\PWV$ (blue squares).  The signature of the
  water vapor absorption on the red end of the $z$-band is apparent in the
  large shift in observed magnitudes that depends on the star color.\label{fig:starchrom}}
\end{figure}

We start with the best estimates of the magnitudes of the calibration stars available from the FGCM fit.
Then define the raw uncalibrated instrumental magnitude for star $j$ observed on exposure $i$,
\begin{equation}
m_b^\inst(i,j) \equiv   -2.5 \log_{10} (\ADU(i,j)) + 2.5 \log_{10}(\Delta T),
\end{equation}
and the retrieval parameter
\begin{equation}
f^\obs(i,j) \equiv 10^{-0.4 \times (m_b^\inst(i,j) - \overline{m_b^\STD(j)}}).
\end{equation}
Eqn. \ref{eqn:mstdfinal} can be used to find,
\begin{equation}
\label{eqn:proxy}
  f^\obs(i,j) = \left(\Inaught^\obs + \Ione^\obs \times \Fprime \right)  \times \left( \frac{\Inaught^\STD}{\Inaught^\STD + \Ione^\STD \times \Fprime} \right).
\end{equation}
Consider the integrals of the observing passbands as unknown in this linear equation, and minimize the sum over calibration stars $j$ on exposure $i$,
\begin{equation}
\label{eqn:proxychi}
\chi^{2}(i) = \sum_j (f^\obs(i,j) - \mathrm{RHS}(\Inaught^\obs, \Ione^\obs))^{2} / \sigma_f^{2},
\end{equation}
where $\mathrm{RHS}$ is the right-hand-side of Eqn. \ref{eqn:proxy}, and $\sigma_f$ is an estimated error for the value of the retrieval parameter $f^\obs$. 
A prescription for evaluation of the appropriate derivatives of the SED is given in Appendix \ref{app:SEDslope}.

The range of colors of the calibration stars on each exposure is generally large enough to project out
reasonable determinations of the chromatic integral $\Ione^\obs$, which we
denote $\mathcal{R}_1$.
The chromatically retrieved $\mathcal{R}_1$ values are compared in Fig.~\ref{fig:i1proxy}
with those computed directly from the FGCM nightly fit parameters.
The ``fudge factors'' given in Appendix \ref{app:SEDslope} were determined to minimize the
differences between the $\Ione$ and $\mathcal{R}_1$ values in these plots.
While the two values being compared in the figure are not independent of each other,
the good agreement confirms that the chromatic corrections are indeed made consistently.
When projected to histograms the differences between the two values are found to be typically $\sim 0.5$ or less,
which corresponds to differences below 5 mmag in the chromatic correction.
While the FGCM fit does not make direct use of the retrieved values, we note
that it includes this information intrinsically as the $\chi^2$
function (Eq.~\ref{eqn:chisq}) is sensitive to the colors of the calibration stars.

\begin{figure}
  \plotone{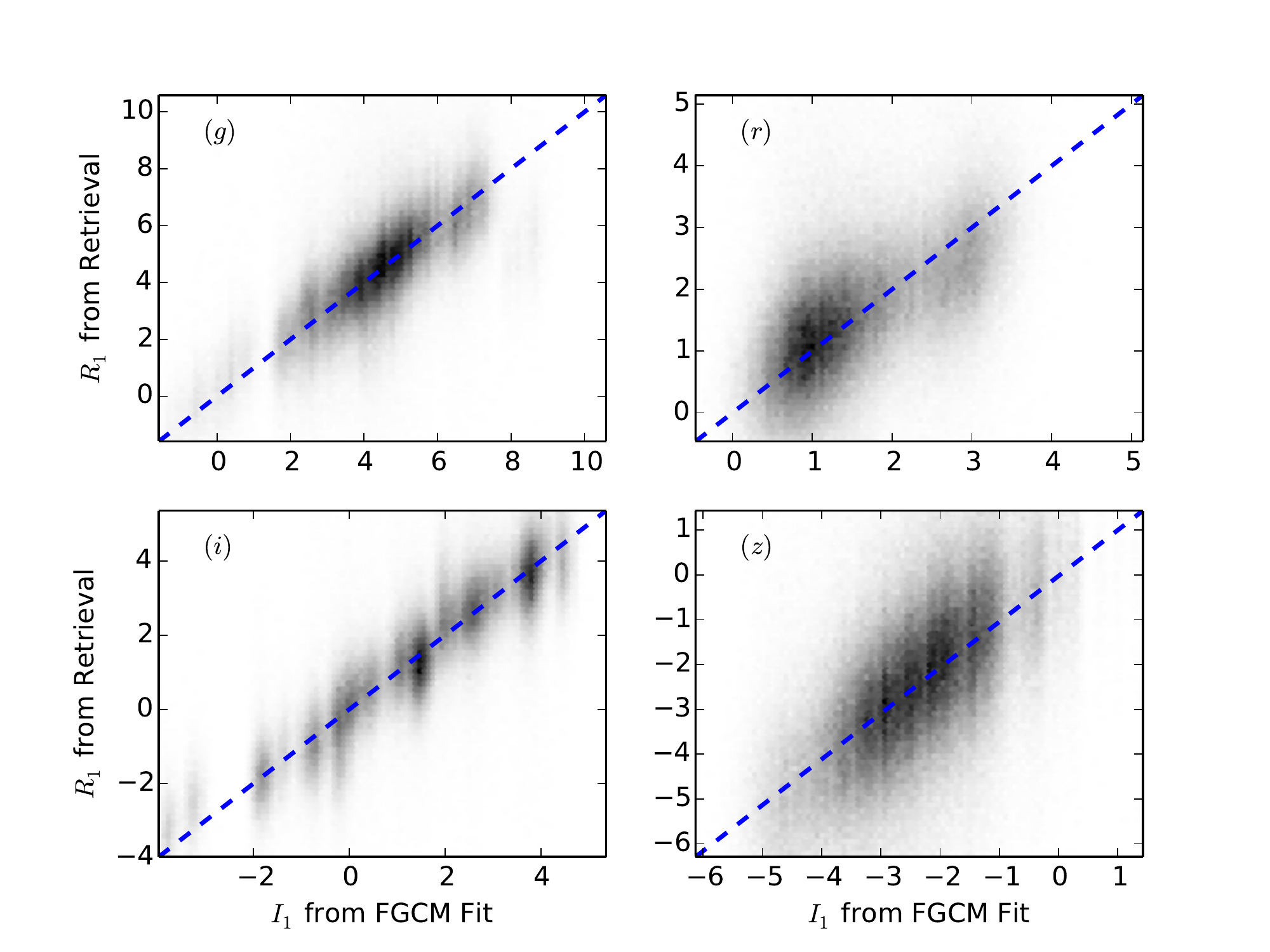}
  \caption{Comparison of $\Ione$ from direct fits and those obtained by
    chromatic retrieval ($\mathcal{R}_1$).}
  \label{fig:i1proxy}
\end{figure}

\section{Initial Preparations}
\label{app:init}

\subsection{FINALCUT Queries}
The DESDM software package accepts raw DECam exposures
and produces ``FINALCUT'' catalogs of observed quantities for each object detected on science exposures.
A set of quality cuts are applied to eliminate exposures on which one or more of a number of recognizable hardware failures
occurred or that were taken through easily detectable cloud cover.
The exposures selected for Y3A1 include wide-field survey and supernova fields, but no standard-star observations.
A control log is then created that drives the calibration process.

Query (SQL) of the DESDM FINALCUT catalogs is done to make an initial selection of observations that
are candidates to be used in the FGCM calibration.
This query requires (Source Extractor data products are indicated in capital letters):
\begin{enumerate}
\item SExtractor flag = 0 (objects that were not deblended, saturated, or had
  other processing problems)
\item The object image is not within 100 pixels of any CCD sensor edge
\item A successful MAG\_PSF fit with $0.001 <$ MAGERR\_PSF $< 0.100\,\mathrm{mag}$  (no explicit cut is made on instrumental magnitude)
\item Selection of stellar sources with CLASS\_STAR $>$ 0.75   \&   -0.003 $<$ SPREAD\_MODEL $<$ 0.003
\end{enumerate}

\subsection{Initial Selection of Calibration Stars}
The following process is used to identify candidate calibration stars: 
\begin{enumerate}
\item Remove observations that were ``blacklisted'' by DESDM due to known instrumental or imaging problems
\item Select $i$-band observations and identify star candidates from detections that are within 1 arcsec of each other
\item Remove candidates that have another candidate within 2 arcsec separation
\item Randomly remove candidates to limit density of stars to approximately one per square arcmin as found at the south Galactic Cap;
HEALPIX (nside = 128) is used in this step
\item Seek observations in the remaining $grzY$ bands that match an $i$-band candidate
\item Identify candidate $griz$ calibration stars as those with at least two observations in each of the four bands
\item Identify the subset of $griz$ calibration stars that also have two observations in $Y$-band
\end{enumerate}
A catalog of $grizY$ FINALCUT observations of candidate calibration stars
is created for use during the FGCM calibration process that follows.

\section{Selections of Calibration Stars, Calibration Exposures, and ``Calibratable'' Nights}
\label{app:build}

On the initial FGCM fit cycle, estimates of the magnitude of each object $j$ in each of the $griz$ bands
are made by computing the observed TOA magnitudes $m_b^\obs(i,j)$ 
(Eqn. \ref{eqn:mtoa2}) with the parameters for the standard atmosphere in Table \ref{table:STDatmo}.
The average value $\overline{m_b^\obs(j)}$ of all observations of that object that
are within 0.10 mag of the brightest is then computed.
If there is not at least a second observation in each band within this tolerance of the brightest,
then the object is removed from the calibration star catalog.
On subsequent cycles calibration stars are required to have been observed on at least two calibration exposures in each of $griz$ bands 
in the previous fit cycle. 

Loose color cuts are applied to eliminate objects far off the stellar locus or simply mismeasured:
\begin{enumerate}
\item $-0.25 < g-r < 2.25$
\item $-0.50 < r-i < 2.25$
\item $-0.50 < i-z < 1.00$
\end{enumerate}

An estimate is made of the ``gray'' extinction of each observation of each calibration star using Eqn.~\ref{eqn:grayext}.
These values are used to choose calibration exposures and calibratable nights.
On the initial FGCM fit cycle, calibration exposures are chosen by requiring that:
\begin{enumerate}
\item There are at least 600 calibration stars visible in the exposure
\item The estimated mean gray extinction of the calibration stars observed on the exposure is less than 0.250 mag 
\item The variance of the gray extinction of the calibration stars observed on the exposure is less than 0.025 mag$^2$
\end{enumerate}
On subsequent cycles the more sophisticated analysis of the individual CCD images
detailed in Section \ref{sec:FGCMfit} is used to define the mean and variance of the gray extinction. 

An observing night is classified ``calibratable'' if:
\begin{enumerate}
\item There were at least 10 calibration exposures on that night
\item The variance of the gray extinction of all calibration exposures on that night is less than 0.100 mag$^2$
\end{enumerate}
There is no requirement that any particular fraction of a night be deemed ``photometric'',
and non-photometric exposures can be taken on a ``calibratable'' night.

Calibration stars for $Y$-band are identified as the subset of $griz$ calibration stars
that also are found on least two $Y$-band calibration exposures.
Calibration $Y$-band exposures are chosen with the same criteria used to select $griz$ calibration exposures,
but the $Y$-band exposures are not used in the FGCM fit.
A separate sequence of analysis cycles is carried out using the $griz$ fit parameters to identify the final set of $Y$-band calibration exposures
used to determine magnitudes for $Y$-band calibration stars.


\end{document}